\documentclass[aip,graphicx]{revtex4-1}

\draft 

\begin{document}

\title{Singular gauge transformations in geometrodynamics} 

\author{Alcides Garat}
\affiliation{1. Former Professor at Universidad de la Rep\'{u}blica, Av. 18 de Julio 1824-1850, 11200 Montevideo, Uruguay.}
\email[]{garat.alcides@gmail.com}
\date{\today}

\begin{abstract}
The new tetrads introduced previously for non-null electromagnetic fields in Einstein-Maxwell spacetimes enable a direct link to the local electromagnetic gauge group of transformations. Due to the peculiar elements in the construction of these new tetrads a direct connection can be established between the local group of electromagnetic gauge transformations and local groups of tetrad transformations on two different local and orthogonal planes of eigenvectors of the Einstein-Maxwell stress-energy tensor. These tetrad vectors are gauge dependent. It is an interesting and relevant problem to study if there are local gauge transformations that can map on the timelike-spacelike plane, the timelike and the spacelike vectors into the intersection of the local light cone and the plane itself. How many of these local gauge transformations exist and how the mathematics and the geometry of these particular transformations play out. These local gauge transformations would be singular and it is important to identify them.
\end{abstract}



\maketitle 

\section{Introduction}
\label{intro}

New tetrads with outstanding properties have been introduced in curved and also flat four-dimensional Lorentzian spacetimes \cite{A,ATGU,MW,IWCP,LomCon}. These tetrad vectors in Einstein-Maxwell spacetimes diagonalize locally and covariantly the stress-energy tensors enabling simplification in evolution algorithms \cite{AEO,SCR,ACD,ENV,AGNS}, for instance. These tetrad vectors establish the connection or link between classical theories of matter and quantum field theories, see references \cite{AYM,gaugeinvmeth,ASU3,ASUN,LE,DSB,PIRT2019,DSBYM}. These tetrad vectors locally determine two orthogonal planes. The plane spanned by the timelike and one of the spacelike vectors is plane or blade one. The orthogonal local plane spanned by the other two spacelike vectors is plane or blade two \cite{SCH}. These new tetrad vectors have in their construction two elements. One of their building blocks is the skeleton. Skeletons are built by using the extremal fields, and extremal fields are found by means of local duality rotations of electromagnetic fields in the Einstein-Maxwell case, for instance. Extremal fields are local electromagnetic gauge invariants in the Abelian electromagnetic case and therefore the skeletons are gauge invariants. The other tetrad construction element is the gauge vector. The gauge vectors are gauge per se, and contain gauge fields in their construction. We can then observe that these tetrad vectors are gauge dependent, locally. A local electromagnetic gauge transformation might be thought of alternatively as a different choice for the gauge vectors, since the gauge vectors are gauge per se, they are a choice, as long as the tetrad vectors do not become trivial. When these new choices are considered and we analyze the change in the tetrad vectors under a local gauge transformation, we see that these vectors transform inside the local planes that they originally defined without leaving them. Keeping in the process the metric tensor invariant. It has been proved through detailed analysis case by case \cite{A} that the local groups of Abelian electromagnetic gauge transformations are mapped into the following local groups of tetrad transformations. On plane one the local group of electromagnetic gauge transformations is mapped to the group $SO(1,1)$ of tetrad boosts, plus two discrete transformations. One of the discrete transformations is the full inversion or just minus the identity in a two by two matrix. The other discrete transformation is called the ``switch'' or flip and we can represent it by a two by two matrix with ones off diagonal and zeroes in the diagonal. It is a reflection and it is not a Lorentz transformation. These set of tetrad transformations make up a new group LB1 or Lorentz blade one group. On plane two the local group of electromagnetic gauge transformations is mapped to $SO(2)$ tetrad spatial rotations which for this specific purpose we called LB2 or Lorentz blade two group. There are local scalars whose four-dimensional gradients through equations found in reference \cite{A} and section III in reference \cite{A} generate electromagnetic gauge transformations that produce tetrad boosts on blade one. These boosts might be composed with full tetrad inversions. These are all proper Lorentz tetrad transformations. There are local scalars that generate boosts composed with the switch, or with the switch and the full inversion. These latter tetrad transformations are special improper. The point of this paper is the following. We set out to investigate if there could be individual, lone gauge tetrad transformations that take in the local plane one, both the timelike and the spacelike vectors into the local light cone that intersects the local plane one. We will investigate this problem for two different geometries. To this goal in section \ref{nullcase} we study the Coulomb case on a flat Minkowskian background. In section \ref{diffeq} we work out the same problem under a more general perspective solving a differential equation for local gauge scalar functions. It is necessary to point out that in manuscript \cite{A} we discussed the vacuum Einstein-Maxwell equations without source terms in the Maxwell equations. Nonetheless we consider important that the discussion of the Coulomb case sets the stage for the discussion of the Reissner-Nordstr\"{o}m case which is a solution to the vacuum Einstein-Maxwell equations in section \ref{diffeqrn}. Finally we will study the general equation in section \ref{geneq}. As a second problem we will study in section \ref{kernelmap} some details of the Kernel of the mapping between the set of electromagnetic local gauge transformations and the LB1 proper sheet. Independently also the Kernel of the mapping between the set of electromagnetic local gauge transformations and LB2=SO(2). The full inversion case and because of its peculiarities will be discussed in detail in section \ref{fullinv}. As a last problem we will prove in sections \ref{switchinvolution}-\ref{mapphomomorph} that LB1 has two sheets, LB1 proper and LB1 special improper. A total of four subsheets. Even though LB1 proper connected to the identity plus the point at infinity is isomorphic to SO(2) as groups, LB1 plus four points at infinity will be homomorphic to SO(2). The points at infinity will result to be the null gauges associated to the tetrad vectors on the local light cones found in sections \ref{nullcase}-\ref{diffeq}-\ref{diffeqrn}-\ref{geneq}. This is a novel result in group theory and we will prove it also using stereographic projections in section \ref{infinity}.

\section{The null case tetrad gauge transformation}
\label{nullcase}

We start by stating that at every point in a curved Lorentzian four-dimensional spacetime there is a duality rotation by an angle $-\alpha$ that transforms a non-null electromagnetic field into an extremal field,

\begin{equation}
\xi_{\mu\nu} = e^{-\ast \alpha} f_{\mu\nu}\ = \cos(\alpha)\:f_{\mu\nu} - \sin(\alpha)\:\ast f_{\mu\nu}.\label{dref}
\end{equation}

where  $\ast f_{\mu\nu}={1 \over 2}\:\epsilon_{\mu\nu\sigma\tau}\:f^{\sigma\tau}$ is the dual tensor of $f_{\mu\nu}$, for the object $\epsilon_{\mu\nu\sigma\tau}$ see section Appendix A in reference \cite{A}. The Levi-Civita pseudotensor can be transformed into a tensor through the use of factors $\sqrt{-g}$, where $g$ is the determinant of the
metric tensor. We use the notation $e_{\alpha\beta\mu\nu}=[\alpha\beta\mu\nu]$ for the covariant components of the Levi-Civita pseudotensor in the Minkowskian frame given in reference \cite{MW}. It is $1$ for an even permutation of $0123$, $-1$ for an odd permutation of $0123$ and $0$ when the indices are not all different. It can be noticed that the signs in $e^{\alpha\beta\mu\nu}$ will be opposite to the standard notation \cite{WE}. The reason for this is that we want to keep the compatibility with the notation in reference \cite{MW} where the definition $e_{0123}=[0123]=1$ was adopted. With these definitions we see that in a spacetime with a metric $g_{\alpha\beta}$,

\begin{equation}
\epsilon^{\alpha\beta\mu\nu}=
{e^{\alpha\beta\mu\nu} \over  \sqrt{-g}}=
- {[\alpha\beta\mu\nu] \over \sqrt{-g}} \ ,\label{lccon}
\end{equation}

are the components of a contravariant tensor \cite{MC2,WE,LL,HS}. The covariant components of (\ref{lccon}) are

\begin{equation}
\epsilon_{\alpha\beta\mu\nu}= e_{\alpha\beta\mu\nu} \sqrt{-g}=
[\alpha\beta\mu\nu] \sqrt{-g} \ ,\label{lccov}
\end{equation}

where

\begin{equation}
g_{\alpha\sigma} g_{\beta\rho}
g_{\mu\kappa} g_{\nu\lambda}\:e^{\sigma\rho\kappa\lambda}= -g\:
e_{\alpha\beta\mu\nu}\ ,
\end{equation}

is satisfied. $F_{\mu\nu}$ is the electromagnetic field and $f_{\mu\nu}= (G^{1/2} / c^2) \: F_{\mu\nu}$ is the geometrized electromagnetic field. The local scalar $\alpha$ is known as the complexion of the electromagnetic field. It is a local gauge invariant quantity. Extremal fields satisfy,

\begin{equation}
\xi_{\mu\nu} \ast \xi^{\mu\nu}= 0\ . \label{i0}
\end{equation}

Equation (\ref{i0}) is a condition imposed on (\ref{dref}) and then the explicit expression for the complexion emerges as $\tan(2\alpha) = - f_{\mu\nu}\:\ast f^{\mu\nu} / f_{\lambda\rho}\:f^{\lambda\rho}$. As antisymmetric fields in a four dimensional Lorentzian spacetime, the extremal fields also verify the identity,

\begin{eqnarray}
\xi_{\mu\alpha}\:\xi^{\nu\alpha} -
\ast \xi_{\mu\alpha}\: \ast \xi^{\nu\alpha} &=& \frac{1}{2}
\: \delta_{\mu}^{\:\:\:\nu}\ Q \ ,\label{i1}
\end{eqnarray}

where $Q=\xi_{\mu\nu}\:\xi^{\mu\nu}=-\sqrt{T_{\mu\nu}T^{\mu\nu}}$ according to equations (39) in \cite{MW}. $Q$ is assumed not to be zero,
because we are dealing with non-null electromagnetic fields. Non-null we clarify means basically that $f_{\mu\nu}\:f^{\mu\nu}\neq0$ and $\ast f_{\mu\nu}\:f^{\mu\nu}\neq0$. In turn and by definitions these last equations imply that $\xi_{\mu\nu}\:\xi^{\mu\nu}\neq0$. The object $T_{\mu\nu}$ is the Einstein-Maxwell stress-energy tensor. It can be proved that condition (\ref{i0}) and through the use of the general identity,

\begin{eqnarray}
A_{\mu\alpha}\:B^{\nu\alpha} -
\ast B_{\mu\alpha}\: \ast A^{\nu\alpha} &=& \frac{1}{2}
\: \delta_{\mu}^{\:\:\:\nu}\: A_{\alpha\beta}\:B^{\alpha\beta}  \ ,\label{ig}
\end{eqnarray}

which is valid for every pair of antisymmetric tensors in a four-dimensional Lorentzian spacetime \cite{MW}, when applied to the case $A_{\mu\alpha} = \xi_{\mu\alpha}$ and $B^{\nu\alpha} = \ast \xi^{\nu\alpha}$ yields the equivalent condition to condition (\ref{i0}),

\begin{eqnarray}
\xi_{\rho\mu}\:\ast \xi^{\mu\nu} &=& 0\ ,\label{i2}
\end{eqnarray}

which is equation (64) in \cite{MW}. The duality rotation given by equation (59) in\cite{MW},

\begin{equation}
f_{\mu\nu} = \xi_{\mu\nu} \: \cos\alpha +
\ast\xi_{\mu\nu} \: \sin\alpha\ ,\label{dr}
\end{equation}

allows us to express the stress-energy tensor in terms of the extremal field,

\begin{equation}
T_{\mu\nu}=\xi_{\mu\lambda}\:\:\xi_{\nu}^{\:\:\:\lambda}
+ \ast \xi_{\mu\lambda}\:\ast \xi_{\nu}^{\:\:\:\lambda}\ .\label{TEMDR}
\end{equation}

Then, we can proceed to introduce the tetrad of eigenvectors to the stress-energy tensor (\ref{TEMDR}),

\begin{eqnarray}
V_{(1)}^{\alpha} &=& \xi^{\alpha\lambda}\:\xi_{\rho\lambda}\:X^{\rho}
\label{V1NONFIXED}\\
V_{(2)}^{\alpha} &=& \sqrt{-Q/2} \:\: \xi^{\alpha\lambda} \: X_{\lambda}
\label{V2NONFIXED}\\
V_{(3)}^{\alpha} &=& \sqrt{-Q/2} \:\: \ast \xi^{\alpha\lambda} \: Y_{\lambda}
\label{V3NONFIXED}\\
V_{(4)}^{\alpha} &=& \ast \xi^{\alpha\lambda}\: \ast \xi_{\rho\lambda}
\:Y^{\rho}\ ,\label{V4NONFIXED}
\end{eqnarray}

where $Q=\xi_{\mu\nu}\:\xi^{\mu\nu}=-\sqrt{T_{\mu\nu}T^{\mu\nu}}$ according to equations (39) in \cite{MW}. After iterative application of equations (\ref{i1}) and (\ref{i2}) we find that the first two (\ref{V1NONFIXED}-\ref{V2NONFIXED}) eigenvectors of the stress-energy tensor have eigenvalue $Q/2$, the last two (\ref{V3NONFIXED}-\ref{V4NONFIXED}) have eigenvalue $-Q/2$. With all these elements it becomes trivial to prove that the tetrad (\ref{V1NONFIXED}-\ref{V4NONFIXED}) is orthogonal and diagonalizes the stress-energy tensor (\ref{TEMDR}). We notice then that we still have to define the vectors $X^{\mu}$ and $Y^{\mu}$. Let us introduce some names. The tetrad vectors have two essential components. For instance in vector $V_{(1)}^{\alpha}$ there are two main structures. First, the skeleton, in this case $\xi^{\alpha\lambda}\:\xi_{\rho\lambda}$, and second, the gauge vector $X^{\rho}$. The gauge vectors it was proved in manuscript \cite{A} could be anything that does not make the tetrad vectors trivial. That is, the tetrad (\ref{V1NONFIXED}-\ref{V4NONFIXED}) diagonalizes the stress-energy tensor for any non-trivial gauge vectors $X^{\mu}$ and $Y^{\mu}$. It was therefore proved that we can make different choices for $X^{\mu}$ and $Y^{\mu}$. In geometrodynamics, the Einstein-Maxwell equations,

\begin{eqnarray}
f^{\mu\nu}_{\:\:\:\:\:;\nu} &=& 0 \label{L1}\\
\ast f^{\mu\nu}_{\:\:\:\:\:;\nu} &=& 0 \label{L2} \\
R_{\mu\nu} &=& f_{\mu\lambda}\:\:f_{\nu}^{\:\:\:\lambda}
+ \ast f_{\mu\lambda}\:\ast f_{\nu}^{\:\:\:\lambda}\ , \label{EM3}
\end{eqnarray}

are telling us that two potential vector fields $A_{\nu}$ and $\ast A_{\nu}$ exist \cite{CF},

\begin{eqnarray}
f_{\mu\nu} &=& A_{\nu ;\mu} - A_{\mu ;\nu}\label{ER} \\
\ast f_{\mu\nu} &=& \ast A_{\nu ;\mu} - \ast A_{\mu ;\nu} \ .\label{DER}
\end{eqnarray}

The symbol $``;''$ stands for covariant derivative with respect to the metric tensor $g_{\mu\nu}$ and the star in $\ast A_{\nu}$ is just a name, not the dual operator, meaning that $\ast A_{\nu ;\mu} = (\ast A_{\nu})_{;\mu}$. The vector fields $A^{\alpha}$ and $\ast A^{\alpha}$ represent a possible choice in geometrodynamics for the vectors $X^{\alpha}$ and $Y^{\alpha}$. It is not meant that the two vector fields have independence from each other, it is just a convenient choice for a particular example. A further justification for the choice $X^{\alpha}=A^{\alpha}$ and $Y^{\alpha}=\ast A^{\alpha}$ could be illustrated through the Reissner-Nordstr\"{o}m geometry. In this particular geometry, $f_{tr}=\xi_{tr}$ and $\ast f_{\theta\phi}=\ast \xi_{\theta\phi}$, therefore, $A_{\theta}=0$ and $A_{\phi}=0$. Then, for the last two tetrad vectors (\ref{V3NONFIXED}-\ref{V4NONFIXED}), the choice $Y^{\alpha}=\ast A^{\alpha}$ becomes meaningful under the light of this particular extreme case, when basically there is no magnetic field. Once we make the choice $X^{\alpha}=A^{\alpha}$ and $Y^{\alpha}=\ast A^{\alpha}$ the question about the geometrical implications of electromagnetic gauge transformations of the tetrad vectors (\ref{V1NONFIXED}-\ref{V4NONFIXED}) arises. We first notice that a local electromagnetic gauge transformation of the ``gauge vectors'' $X^{\alpha}=A^{\alpha}$ and $Y^{\alpha}=\ast A^{\alpha}$ can be just interpreted as a new choice for the gauge vectors $X_{\alpha} = A_{\alpha} + \Lambda_{,\alpha}$ and $Y_{\alpha} = \ast A_{\alpha} + \ast \Lambda_{,\alpha}$. When we make the transformation, $A_{\alpha} \rightarrow A_{\alpha} + \Lambda_{,\alpha}$, $f_{\mu\nu}$ remains invariant, and the transformation, $\ast A_{\alpha} \rightarrow \ast A_{\alpha} + \ast \Lambda_{,\alpha}$, leaves $\ast f_{\mu\nu}$ invariant, as long as the functions $\Lambda$ and $\ast \Lambda$ are scalars. It is valid to ask how the tetrad vectors (\ref{V1NONFIXED}-\ref{V2NONFIXED}) will transform under $A_{\alpha} \rightarrow A_{\alpha} + \Lambda_{,\alpha}$ and (\ref{V3NONFIXED}-\ref{V4NONFIXED}) under $\ast A_{\alpha} \rightarrow \ast A_{\alpha} + \ast \Lambda_{,\alpha}$.

For example, from reference \cite{A} a particular boost after the gauge transformation would look like,

\begin{eqnarray}
{\tilde{V}_{(1)}^{\alpha}
\over \sqrt{-\tilde{V}_{(1)}^{\beta}\:\tilde{V}_{(1)\beta}}}&=&
{(1+C) \over \sqrt{(1+C)^2-D^2}}
\:{V_{(1)}^{\alpha} \over \sqrt{-V_{(1)}^{\beta}\:V_{(1)\beta}}}+
{D \over \sqrt{(1+C)^2-D^2}}
\:{V_{(2)}^{\alpha} \over \sqrt{V_{(2)}^{\beta}\:V_{(2)\beta}}}\label{TN1N}\\
{\tilde{V}_{(2)}^{\alpha}
\over \sqrt{\tilde{V}_{(2)}^{\beta}\:\tilde{V}_{(2)\beta}}}&=&
{D \over \sqrt{(1+C)^2-D^2}}
\:{V_{(1)}^{\alpha} \over \sqrt{-V_{(1)}^{\beta}\:V_{(1)\beta}}} +
{(1+C) \over \sqrt{(1+C)^2-D^2}}
\:{V_{(2)}^{\alpha} \over \sqrt{V_{(2)}^{\beta}\:V_{(2)\beta}}}\ .
\label{TN2N}
\end{eqnarray}

In equations (\ref{TN1N}-\ref{TN2N}) the following notation has been used, $C = (-Q/2)\:V_{(1)\sigma}\:\Lambda^{\sigma} / (\:V_{(2)\beta}\:V_{(2)}^{\beta}\:)$, $D = (-Q/2)\:V_{(2)\sigma}\:\Lambda^{\sigma} / (\:V_{(1)\beta}\:V_{(1)}^{\beta}\:)$ and $[(1+C)^2-D^2]>0$ must be satisfied. The notation $\Lambda^{\alpha}$ has been used for $\Lambda^{,\alpha}$ where $\Lambda$ is the local scalar generating the local gauge transformation. $U^{\alpha} = {V_{(1)}^{\alpha} \over \sqrt{-V_{(1)}^{\beta}\:V_{(1)\beta}}}$ and $V^{\alpha} = {V_{(2)}^{\alpha} \over \sqrt{V_{(2)}^{\beta}\:V_{(2)\beta}}}$ according to the notation used in paper \cite{A},

\begin{eqnarray}
V_{(1)}^{\alpha} &=& \xi^{\alpha\lambda}\:\xi_{\rho\lambda}\:A^{\rho}
\label{V1}\\
V_{(2)}^{\alpha} &=& \sqrt{-Q/2} \: \xi^{\alpha\lambda} \: A_{\lambda}
\label{V2}\\
V_{(3)}^{\alpha} &=& \sqrt{-Q/2} \: \ast \xi^{\alpha\lambda}
\: \ast A_{\lambda}\label{V3}\\
V_{(4)}^{\alpha} &=& \ast \xi^{\alpha\lambda}\: \ast \xi_{\rho\lambda}
\:\ast A^{\rho}\ .\label{V4}
\end{eqnarray}

$Q$ is assumed not to be zero, because we are dealing with non-null electromagnetic fields. For the particular case when $1+C > 0$, the transformations (\ref{TN1N}-\ref{TN2N}) manifest that an electromagnetic gauge transformation on the vector field $A^{\alpha} \rightarrow A^{\alpha} + \Lambda^{\alpha}$, that leaves invariant the electromagnetic field $f_{\mu\nu}$, generates a boost transformation on the normalized tetrad vector fields $\left({V_{(1)}^{\alpha} \over \sqrt{-V_{(1)}^{\beta}\:V_{(1)\beta}}}, {V_{(2)}^{\alpha} \over \sqrt{V_{(2)}^{\beta}\:V_{(2)\beta}}}\right)$. In this case $\cosh(\phi) = {(1+C) \over \sqrt{(1+C)^2-D^2}}$. This was just one of the possible cases in LB1. Similar analysis for the vector gauge transformations in the local plane two generated by ($Z^{\alpha}, W^{\alpha}$). See reference \cite{A} section III for the detailed analysis of all possible cases. In essence the local group of electromagnetic Abelian gauge transformations is proved to be mapped into the local group LB1 of local tetrad transformations on plane one as discussed in the introduction. On the local plane two the local group of electromagnetic Abelian gauge transformations is proved to be mapped into the local group LB2=$SO(2)$ of local tetrad transformations as discussed in the introduction as well, see also reference \cite{A} for the details. Schouten defined what he called, a two-bladed structure in a spacetime \cite{SCH}. These local blades or planes are the planes determined by the pairs ($V_{(1)}^{\alpha}, V_{(2)}^{\alpha}$) and ($V_{(3)}^{\alpha}, V_{(4)}^{\alpha}$). Even though in our original paper \cite{A} we are dealing with vacuum Maxwell equations without source terms we will proceed to analyze the Coulomb case, which shares similarities in gauge analysis with the Reissner-Nordstr\"{o}m case which is a solution to the vacuum Einstein-Maxwell equations, for instance on the local plane one. The results in section \ref{nullcase} are also valid for the Maxwell equations with sources $J^{\mu}$ in Minkowski spacetime were the particular tetrad construction and gauge analysis are presented in section \ref{sec:appI}.


Let us start with the case where the gauge choice is $f_{tr} = e/r^{2}$, $A_{t} = e/r$ and $A_{r} = 0$. Let us analyze the components of the tetrad vectors (\ref{V1}-\ref{V2}) for this case for a flat Minkowskian spacetime with signature $(-+++)$.

\begin{eqnarray}
V_{(1)}^{t} &=& \xi^{tr}\:\xi_{tr}\:A^{t} = \mid \xi_{tr} \mid^{2}\:A_{t} \label{V1tEX}\\
V_{(1)}^{r} &=& \xi^{rt}\:\xi_{rt}\:A^{r} = 0 \label{V1rEX}\\
V_{(2)}^{t} &=& \mid \xi_{tr} \mid \: \xi^{tr} \: A_{r} = 0 \label{V2tEX}\\
V_{(2)}^{r} &=& \mid \xi_{tr} \mid \: \xi^{rt} \: A_{t} = \mid \xi_{tr} \mid\:\xi_{tr}\:A_{t} \ .\label{V2rEX}
\end{eqnarray}

where $Q = -2\mid \xi_{tr} \mid^{2}$. Next let us proceed to analyze the norm of these different orthogonal vectors.

\begin{eqnarray}
\lefteqn{ V_{(1)}^{\alpha}\:V_{(1)\alpha} = V_{(1)}^{t}\:V_{(1)t} + V_{(1)}^{r}\:V_{(1)r} }\nonumber \\ &&= -\mid \xi_{tr} \mid^{4}\:(A_{t})^{2} +
\mid \xi_{tr} \mid^{4}\:(A_{r})^{2} = -\mid \xi_{tr} \mid^{4}\:(A_{t})^{2} \label{FPCB}\\
&& V_{(2)}^{\alpha}\:V_{(2)\alpha} = V_{(2)}^{t}\:V_{(2)t} + V_{(2)}^{r}\:V_{(2)r}  \nonumber \\ &&= -\mid \xi_{tr} \mid^{4}\:(A_{r})^{2} +
\mid \xi_{tr} \mid^{4}\:(A_{t})^{2} = \mid \xi_{tr} \mid^{4}\:(A_{t})^{2} \ .\label{SPCB}
\end{eqnarray}

where the relation $V_{(1)}^{\alpha}\:V_{(1)\alpha} = -V_{(2)}^{\alpha}\:V_{(2)\alpha} \ne 0$ is evident.

We then proceed immediately to the following particular case. The Coulomb example where $f_{tr} = e/r^{2} = \xi_{tr}$, $A_{t} = e/r$ and $A^{new}_{r} = -e/r$. If we do not write ```new'' we mean the original components before the local electromagnetic gauge transformation. We reiterate that all the results in section \ref{nullcase} are also valid for the Maxwell equations with sources $J^{\mu}$ in Minkowski spacetime were the particular tetrad construction and gauge analysis are presented in section \ref{sec:appI}. The analysis in section \ref{nullcase} is also valid for Einstein-Maxwell equations in curved spacetimes with sources $J^{\mu}$ were the particular tetrad construction and gauge analysis are completely analogous to the discussion presented in section \ref{sec:appI}. Resuming our discussion this Coulomb case corresponds to a gauge transformation $\Lambda_{t} = 0$ and $\Lambda_{r} = -e/r$ of the original gauge choice $A_{t} = e/r$ and $A_{r} = 0$. We notice that a local electromagnetic gauge transformation of the ``gauge vectors'' $X^{\alpha}=A^{\alpha}$ and $Y^{\alpha}$ (see section \ref{sec:appI}) can be just interpreted as a new choice for the gauge vectors $X_{\alpha} = A_{\alpha} + \Lambda_{,\alpha}$ and $Y_{\alpha} \rightarrow Y_{\alpha} + \ast \Lambda_{,\alpha}$. For simplicity we will use the notation for local gauge transformations $\Lambda_{,\mu} = \Lambda_{\mu}$ where $\Lambda$ is a local scalar. Let us analyze the components of the tetrad vectors (\ref{V1}-\ref{V2}) for the new case raised for a flat Minkowskian spacetime with signature $(-+++)$.

\begin{eqnarray}
V_{(1)}^{t} &=& \xi^{tr}\:\xi_{tr}\:A^{t} = \mid \xi_{tr} \mid^{2}\:A_{t} \label{V1tEX}\\
V_{(1)}^{r} &=& \xi^{rt}\:\xi_{rt}\:A_{new}^{r} = -\mid \xi_{tr} \mid^{2}\:A^{new}_{r} \label{V1rEX}\\
V_{(2)}^{t} &=& \mid \xi_{tr} \mid \: \xi^{tr} \: A^{new}_{r} = -\mid \xi_{tr} \mid\:\xi_{tr}\:A^{new}_{r} \label{V2tEX}\\
V_{(2)}^{r} &=& \mid \xi_{tr} \mid \: \xi^{rt} \: A_{t} = \mid \xi_{tr} \mid\:\xi_{tr}\:A_{t} \ .\label{V2rEX}
\end{eqnarray}

where $Q = -2\mid \xi_{tr} \mid^{2}$. Next let us proceed to analyze the norm of these vectors.

\begin{eqnarray}
\lefteqn{ V_{(1)}^{\alpha}\:V_{(1)\alpha} = V_{(1)}^{t}\:V_{(1)t} + V_{(1)}^{r}\:V_{(1)r} }\nonumber \\ &&= -\mid \xi_{tr} \mid^{4}\:(A_{t})^{2} +
\mid \xi_{tr} \mid^{4}\:(A^{new}_{r})^{2} = \mid \xi_{tr} \mid^{4}\:(-(A_{t})^{2} + (A^{new}_{r})^{2}) \label{FPC}\\
&& V_{(2)}^{\alpha}\:V_{(2)\alpha} = V_{(2)}^{t}\:V_{(2)t} + V_{(2)}^{r}\:V_{(2)r}  \nonumber \\ &&= -\mid \xi_{tr} \mid^{4}\:(A^{new}_{r})^{2} +
\mid \xi_{tr} \mid^{4}\:(A_{t})^{2} = \mid \xi_{tr} \mid^{4}\:((A_{t})^{2} - (A^{new}_{r})^{2}) \ .\label{SPC}
\end{eqnarray}

where the relation $V_{(1)}^{\alpha}\:V_{(1)\alpha} = -V_{(2)}^{\alpha}\:V_{(2)\alpha}$ is evident. From the detailed analysis reproduced from reference \cite{A} in section IV ``gauge geometry'' we calculate the coefficients C and D from equations (54-55). We then will identify the coefficients in equations (\ref{FPC}-\ref{SPC}) for the Coulomb case with the coefficients in the general equations (56-57) from reference \cite{A} given by $\tilde{V}_{(1)}^{\alpha}\:\tilde{V}_{(1)\alpha} = [(1+C)^2-D^2]\:V_{(1)}^{\alpha}\:V_{(1)\alpha}$ and $\tilde{V}_{(2)}^{\alpha}\:\tilde{V}_{(2)\alpha} =
[(1+C)^2-D^2]\:V_{(2)}^{\alpha}\:V_{(2)\alpha}$.

\begin{eqnarray}
C&=&(-Q/2)\:V_{(1)\sigma}\:\Lambda^{\sigma} / (\:V_{(2)\beta}\:
V_{(2)}^{\beta}\:) = \mid \xi_{tr} \mid^{2}\:{(\Lambda^{t}\:V_{(1)t} + \Lambda^{r}\:V_{(1)r}) \over \mid \xi_{tr} \mid^{4}\:((A_{t})^{2} + (A_{r})^{2})}\label{COEFFCAT1}\\
D&=&(-Q/2)\:V_{(2)\sigma}\:\Lambda^{\sigma} / (\:V_{(1)\beta}\:
V_{(1)}^{\beta}\:) = \mid \xi_{tr} \mid^{2}\:{(\Lambda^{t}\:V_{(2)t} + \Lambda^{r}\:V_{(2)r}) \over \mid \xi_{tr} \mid^{4}\:(-(A_{t})^{2} + (A_{r})^{2})}\ .\label{COEFFDAT2}
\end{eqnarray}

Both these coefficients (\ref{COEFFCAT1}-\ref{COEFFDAT2}) after some simple algebra reduce to,

\begin{eqnarray}
C&=& -{(\Lambda^{t}\:A_{t} + \Lambda^{r}\:A_{r}) \over ((A_{t})^{2} + (A_{r})^{2})}\label{COEFFCBT1}\\
D&=& {\xi_{tr} \over \mid \xi_{tr} \mid}\:{(\Lambda^{t}\:A_{r} + \Lambda^{r}\:A_{t}) \over (-(A_{t})^{2} + (A_{r})^{2})}\ .\label{COEFFDBT2}
\end{eqnarray}

In the last general equation (\ref{COEFFCBT1}) we notice the following. The original gauge from which we transform is $A_{t} = e/r$ and $A_{r} = 0$. The local gauge transformation is $\Lambda_{t} = 0$ and $\Lambda_{r} = -e/r$. Therefore, the $A_{r}$ in the coefficient C is zero, because it is the old $A_{r}$. Therefore, $C = 0$. For the second equation (\ref{COEFFDBT2}) knowing that $\xi_{tr} = f_{tr} = e/r^{2}$, we find $D =-{\Lambda^{r} \over A_{t}} = -{\Lambda_{r} \over A_{t}}$.
Finally, we calculate the norm transformation coefficient in equations (56-57) of manuscript \cite{A}.

\begin{eqnarray}
[(1+C)^2-D^2] = (1-0)^{2} - ({\Lambda_{r} \over A_{t}})^{2} = {A_{t}^{2} - \Lambda_{r}^{2} \over A_{t}^{2}} = 0 \label{NULLCASE}
\end{eqnarray}

The same result can be noticed from the particular analysis in equations (\ref{FPC}-\ref{SPC}). Because the new gauge is a very special gauge for which $A_{t} = - \Lambda_{r} = e/r$. We will prove in the next section that it is the inhomogeneous solution to a differential equation. For this particular gauge transformation both original vectors (\ref{V1}-\ref{V2}) one timelike and the other spacelike are transformed into null vectors on the local light cone. It is a singular and unique gauge transformation. Only one in an infinite set. It is in fact a set of measure zero in the whole set of gauge transformations.

\section{Gauge differential equation: Coulomb case}
\label{diffeq}

All the results in section \ref{nullcase} are also valid for the Maxwell equations with sources $J^{\mu}$ in Minkowski spacetime were the particular tetrad construction and gauge analysis are presented in section \ref{sec:appI}. In this section we will proceed following an inverse path. We will impose the null condition for gauge vector transformation and from the ensuing differential equation on the local scalar we will find the local gauge transformations that take a timelike and a spacelike vectors into the same null vector on the local light cone. Let us then impose the condition $D = 1+ C$ in accordance to the general theory of tetrad gauge transformations through equations (56-57) in reference \cite{A}. We briefly remind ourselves that two cases are possible for $[(1+C)^2-D^2]=0$ and the case $D = -(1+ C)$ is analyzed in a similar way leading to similar results. Next we transform from $A_{t} = e/r$ and $A_{r} = 0$. We will not include hereafter the ```new'' label because we do not want to overload with notation and because it is easy to follow the components before and after the local gauge transformations. The general local gauge transformation is in principle $\Lambda_{t} \ne 0$ and $\Lambda_{r} \ne 0$. Let us use the general equations (\ref{COEFFCBT1}-\ref{COEFFDBT2}) for the coefficients C and D in the equation $D = 1+ C$,

\begin{eqnarray}
{\xi_{tr} \over \mid \xi_{tr} \mid}\:{(\Lambda^{t}\:A_{r} + \Lambda^{r}\:A_{t}) \over (-(A_{t})^{2} + (A_{r})^{2})} = 1 + (-){(\Lambda^{t}\:A_{t} + \Lambda^{r}\:A_{r}) \over ((A_{t})^{2} + (A_{r})^{2})} \label{generalD+C}
\end{eqnarray}

We start from $A_{t} = e/r$ and $A_{r} = 0$ and $\xi_{tr} = f_{tr} = e/r^{2}$. Therefore, we are left with,

\begin{eqnarray}
-{\Lambda^{r} \over A_{t}} = 1 - {\Lambda^{t} \over A_{t}} \label{generalD+Cparticular}
\end{eqnarray}

We can rewrite this equation as,

\begin{eqnarray}
\Lambda^{t} - \Lambda^{r} = A_{t} \ , \label{gaugediffeq}
\end{eqnarray}

which is equivalent to,

\begin{eqnarray}
-\Lambda_{t} - \Lambda_{r} = A_{t} \ , \label{gaugediffeq2}
\end{eqnarray}

Taking cross derivatives with respect to coordinates $t$ and $r$ and reminding about the integrability condition $\Lambda_{tr} = \Lambda_{rt}$ and after some simple algebra knowing that $\partial_{t}A_{t} = 0$ we find,

\begin{eqnarray}
\Lambda_{tt} - \Lambda_{rr} = \partial_{r}A_{t} \ . \label{gaugediffeq3}
\end{eqnarray}

The inhomogeneous solution is $\Lambda_{r} = -e/r = -A_{t}$. The homogeneous solution can be found to be $\Lambda_{H} = A\:\cos\omega(t-r) + B\:\sin\omega(t-r)$ which are gauge waves traveling to the future at the speed of light where $A$, $B$ and $\omega$ are constants. We are contemplating not only equation (\ref{gaugediffeq3}) but also (\ref{gaugediffeq2}). It is just one inhomogeneous gauge transformation, therefore a set of measure zero. The inhomogeneous solution for $D=-(1+C)$ in the past light cone will correspond to $\Lambda_{r} = e/r = A_{t}$. Reflections about the light cone leave the solutions already found invariant.

\section{Gauge differential equation: Reissner-Nordstr\"{o}m case}
\label{diffeqrn}

The line element for this spacetime is given by the following expression \cite{RW,MC2,monopole},

\begin{eqnarray}
ds^{2} = - (1 - {2m \over r} + {e^{2} \over r^{2}})\: dt^{2} + (1 - {2m \over r} + {e^{2} \over r^{2}})^{-1}\: dr^{2} + r^{2}\:(d\theta^{2} + \sin^{2}\theta\:d\phi^{2})\ . \label{reissnord}
\end{eqnarray}

In this section we will proceed following the same path as in section \ref{diffeq} for the Coulomb case. Once again we will impose the null condition for gauge vector transformation and from the ensuing differential equation on the local scalar we will find the local gauge transformations that take a timelike and a spacelike vectors into the same null vector on the local light cone. This time on a solution to the vacuum Einstein-Maxwell equations without sources in the Maxwell equations. Let us then impose the condition $D = 1+ C$ in accordance to the general theory of tetrad gauge transformations through equations (56-57) in reference \cite{A}. Next we transform from $A_{t} = e/r$ and $A_{r} = 0$. The general local gauge transformation is in principle $\Lambda_{t} \ne 0$ and $\Lambda_{r} \ne 0$. Let us analyze the components of the tetrad vectors (\ref{V1}-\ref{V2}) for the Reissner-Nordstr\"{o}m case with signature $(-+++)$.

\begin{eqnarray}
V_{(1)}^{t} &=& \xi^{tr}\:\xi_{tr}\:A^{t} = -g^{tt}\mid \xi_{tr} \mid^{2}\:A_{t} \label{V1tRN}\\
V_{(1)}^{r} &=& \xi^{rt}\:\xi_{rt}\:A^{r} = -g^{rr}\mid \xi_{tr} \mid^{2}\:A_{r} \label{V1rRN}\\
V_{(2)}^{t} &=& \mid \xi_{tr} \mid \: \xi^{tr} \: A_{r} = -\mid \xi_{tr} \mid\:\xi_{tr}\:A_{r} \label{V2tRN}\\
V_{(2)}^{r} &=& \mid \xi_{tr} \mid \: \xi^{rt} \: A_{t} = \mid \xi_{tr} \mid\:\xi_{tr}\:A_{t} \ .\label{V2rRN }
\end{eqnarray}

where $Q = -2\mid \xi_{tr} \mid^{2}$. Next let us proceed to analyze the norm of these vectors.

\begin{eqnarray}
\lefteqn{ V_{(1)}^{\alpha}\:V_{(2)\alpha} = V_{(1)}^{t}\:V_{(2)t} + V_{(1)}^{r}\:V_{(2)r} }\nonumber \\ &&= g^{tt}\:g_{tt}\:\mid \xi_{tr} \mid^{3}\:\xi_{tr}\:A_{t}\:A_{r} -
g^{rr}\:g_{rr}\:\mid \xi_{tr} \mid^{3}\:\xi_{tr}\:A_{r}\:A_{t} \nonumber \\ &&= \mid \xi_{tr} \mid^{3}\:\xi_{tr}\:A_{t}\:A_{r}\:(g^{tt}\:g_{tt}-g^{rr}\:g_{rr}) = 0 \label{FSPCRN}\\
\lefteqn{ V_{(1)}^{\alpha}\:V_{(1)\alpha} = V_{(1)}^{t}\:V_{(1)t} + V_{(1)}^{r}\:V_{(1)r} }\nonumber \\ &&= \mid \xi_{tr} \mid^{4}\:(g^{tt}\:(A_{t})^{2} +
g^{rr}\:(A_{r})^{2}) = -\mid \xi_{tr} \mid^{4}\:(g_{rr}\:(A_{t})^{2} + g_{tt}\:(A_{r})^{2}) \label{FPCRN}\\
&& V_{(2)}^{\alpha}\:V_{(2)\alpha} = V_{(2)}^{t}\:V_{(2)t} + V_{(2)}^{r}\:V_{(2)r} = \mid \xi_{tr} \mid^{4}\:(g_{rr}\:(A_{t})^{2} + g_{tt}\:(A_{r})^{2}) \ .\label{SPCRN}
\end{eqnarray}

where the relation $V_{(1)}^{\alpha}\:V_{(1)\alpha} = -V_{(2)}^{\alpha}\:V_{(2)\alpha}$ is evident. From the detailed analysis reproduced from reference \cite{A} in section IV ``gauge geometry'' we calculate the coefficients C and D from equations (54-55). We then will identify the coefficients in equations (\ref{FPCRN}-\ref{SPCRN}) for the Reissner-Nordstr\"{o}m case with the coefficients in the general equations (56-57) in reference \cite{A} $\tilde{V}_{(1)}^{\alpha}\:\tilde{V}_{(1)\alpha} = [(1+C)^2-D^2]\:V_{(1)}^{\alpha}\:V_{(1)\alpha}$ and $\tilde{V}_{(2)}^{\alpha}\:\tilde{V}_{(2)\alpha} = [(1+C)^2-D^2]\:V_{(2)}^{\alpha}\:V_{(2)\alpha}$.

\begin{eqnarray}
C&=&(-Q/2)\:V_{(1)\sigma}\:\Lambda^{\sigma} / (\:V_{(2)\beta}\:
V_{(2)}^{\beta}\:) = \mid \xi_{tr} \mid^{2}\:{(\Lambda^{t}\:V_{(1)t} + \Lambda^{r}\:V_{(1)r}) \over \mid \xi_{tr} \mid^{4}\:(g_{rr}\:(A_{t})^{2} + g_{tt}\:(A_{r})^{2})}\label{COEFFCAT1RN}\\
D&=&(-Q/2)\:V_{(2)\sigma}\:\Lambda^{\sigma} / (\:V_{(1)\beta}\:
V_{(1)}^{\beta}\:) = -\mid \xi_{tr} \mid^{2}\:{(\Lambda^{t}\:V_{(2)t} + \Lambda^{r}\:V_{(2)r}) \over \mid \xi_{tr} \mid^{4}\:(g_{rr}\:(A_{t})^{2} + g_{tt}\:(A_{r})^{2})}\ .\label{COEFFDAT2RN}
\end{eqnarray}

Both these coefficients (\ref{COEFFCAT1RN}-\ref{COEFFDAT2RN}) after some simple algebra reduce to,

\begin{eqnarray}
C&=&  -{(\Lambda^{t}\:A_{t} + \Lambda^{r}\:A_{r}) \over (g_{rr}\:(A_{t})^{2} + g_{tt}\:(A_{r})^{2})}\label{COEFFCBT1RN}\\
D&=&  -{\xi_{tr} \over \mid \xi_{tr} \mid}\:{(-g_{tt}\:\Lambda^{t}\:A_{r} + g_{rr}\:\Lambda^{r}\:A_{t}) \over (g_{rr}\:(A_{t})^{2} + g_{tt}\:(A_{r})^{2})}\ .\label{COEFFDBT2RN}
\end{eqnarray}

Let us then impose the singular condition $D = 1+C$,

\begin{eqnarray}
-{\xi_{tr} \over \mid \xi_{tr} \mid}\:{(-g_{tt}\:\Lambda^{t}\:A_{r} + g_{rr}\:\Lambda^{r}\:A_{t}) \over (g_{rr}\:(A_{t})^{2} + g_{tt}\:(A_{r})^{2})}\  = 1 + (-){(\Lambda^{t}\:A_{t} + \Lambda^{r}\:A_{r}) \over (g_{rr}\:(A_{t})^{2} + g_{tt}\:(A_{r})^{2})}\label{generalD+CRN}
\end{eqnarray}

Let us consider the original gauge as $A_{t} = e/r$ and $A_{r} = 0$ and $\xi_{tr} = f_{tr} = e/r^{2}$, knowing that $\Lambda^{t} = g^{tt}\:\Lambda_{t}$ and $\Lambda^{r} = g^{rr}\:\Lambda_{r}$. Therefore, we are left with,

\begin{eqnarray}
{g_{tt}\:\Lambda_{r} \over A_{t}} = 1 + {\Lambda_{t} \over A_{t}} \label{generalD+CparticularRN}
\end{eqnarray}

We can rewrite this equation as,

\begin{eqnarray}
\Lambda_{t} - g_{tt}\:\Lambda_{r} = -A_{t} \ , \label{gaugediffeqRN}
\end{eqnarray}



Taking cross derivatives with respect to coordinates $t$ and $r$ and reminding about the integrability condition $\Lambda_{tr} = \Lambda_{rt}$ and after some simple algebra knowing that $\partial_{t}A_{t} = 0$ we find,

\begin{eqnarray}
\Lambda_{tt} - g_{tt}\:\partial_{r}g_{tt} \Lambda_{r} - (g_{tt})^{2} \Lambda_{rr}= -g_{tt}\:\partial_{r}A_{t} \ . \label{gaugediffeq3RN}
\end{eqnarray}

The inhomogeneous solution is $\Lambda_{r} = -g_{rr}\:A_{t}$. The homogeneous solution can be found to be a linear combination of the real and imaginary parts of,

\begin{eqnarray}
\Lambda_{H} = \exp(-\imath\int_{a}^{r}({\omega \over g_{tt}} + k)\:dr)\:\exp\imath(k\:r-\omega\:t) \ , \label{INHRN}
\end{eqnarray}

which are gauge waves with $a$, $k$ and $\omega$ constants. We are contemplating not only equation (\ref{gaugediffeq3RN}) but also (\ref{gaugediffeqRN}). There is only a unique inhomogeneous solution $\Lambda_{r} = -{e \over r\:(1 - {2m \over r} + {e^{2} \over r^{2}})} = A_{t}/g_{tt}$. It is just one inhomogeneous gauge transformation, therefore a set of measure zero. The inhomogeneous solution for $D=-(1+C)$ in the past light cone will correspond to $\Lambda_{r} = g_{rr}\:A_{t}$. Reflections about the light cone leave the solutions already found invariant.

\section{General differential equation}
\label{geneq}

In this final section we will present the general differential equation from which all the particular problems previously studied arise.

\begin{eqnarray}
C&=&(-Q/2)\:V_{(1)\sigma}\:\Lambda^{\sigma} / (\:V_{(2)\beta}\:
V_{(2)}^{\beta}\:) \label{COEFFCGEN}\\
D&=&(-Q/2)\:V_{(2)\sigma}\:\Lambda^{\sigma} / (\:V_{(1)\beta}\:
V_{(1)}^{\beta}\:) \ .\label{COEFFDGEN}
\end{eqnarray}

where the relation $V_{(1)}^{\alpha}\:V_{(1)\alpha} = -V_{(2)}^{\alpha}\:V_{(2)\alpha}$ has been used. When we impose the condition $D = 1+ C$ we obtain the following differential equation on the local scalar gradient $\Lambda^{\sigma}$.

\begin{eqnarray}
(-Q/2)\:V_{(2)\sigma}\:\Lambda^{\sigma} / (\:V_{(1)\beta}\:
V_{(1)}^{\beta}\:) = 1 + (-Q/2)\:V_{(1)\sigma}\:\Lambda^{\sigma} / (\:V_{(2)\beta}\:
V_{(2)}^{\beta}\:) \ . \label{DIFFEQGEN}
\end{eqnarray}

By multiplying both sides by $V_{(1)}^{\alpha}\:V_{(1)\alpha} = -V_{(2)}^{\alpha}\:V_{(2)\alpha}$ we get,

\begin{eqnarray}
(-Q/2)\:V_{(2)\sigma}\:\Lambda^{\sigma}  =
V_{(1)\beta}\:V_{(1)}^{\beta} - (-Q/2)\:V_{(1)\sigma}\:\Lambda^{\sigma}  \ . \label{DIFFEQGEN1}
\end{eqnarray}

This is a differential equation on the gradient $\Lambda^{\sigma}$ with a source term $V_{(1)\beta}\:V_{(1)}^{\beta}$. Therefore it will possess an inhomogeneous solution and homogeneous solutions. Let us not forget that for the case $D = -(1+C)$ there is also one more singular inhomogeneous solution causing the timelike and spacelike vectors on the local plane one to transform into null vectors on the local light cone. This additional solution $D = -(1+C)$ corresponds to an inhomogeneous solution in the past light cone. Reflections about the light cone leave the solutions already found invariant.

\section{Kernel of the mapping}
\label{kernelmap}

The proof to the following theorems can be found in references \cite{A} and \cite{ROMP}. In order to summarize all the results in this section which we will need in the subsequent sections we state the following,

\begin{itemize}

\item In reference \cite{ROMP} it was found that the Kernel of the map between the local group of electromagnetic gauge transformations and the local group of tetrad transformations in the proper sector in the local blade one is just composed by the group $PGB2$, where $PGB2=\{\Lambda / \Lambda^{\mu} \in \mbox{local Plane 2} \}$ is the set of pure gauge in blade two, as long as the choice for tetrad gauge vector is not pure gauge $X^{\mu} \neq \Lambda^{\mu}$ or the pure gauge multiplied by a local scalar $X^{\mu} \neq \frac{1}{C}\:\Lambda^{\mu}$ with $1+C>0$ which is equivalent to pure gauge for the skeleton-gauge vector tetrad structure on plane one. The group $PGB2$ is of measure zero and it is just gradients of scalars in a local plane in a four-dimensional spacetime. There is an isomorphism between the group of local electromagnetic gauge transformations minus the set $PGB2$ and LB1.

\item In reference \cite{ROMP} it was found that the Kernel of the map between the local group of electromagnetic gauge transformations and the local group of tetrad transformations on the local blade two is just composed by the group $PGB1$, where $PGB1=\{\Lambda / \Lambda^{\mu} \in \mbox{local Plane 1} \}$ is the set of pure gauge in blade one, as long as the choice for tetrad gauge vector is not pure gauge $Y^{\mu} \neq \ast \Lambda^{\mu}$ or the pure gauge multiplied by a local scalar $X^{\mu} \neq \frac{1}{N}\:\ast \Lambda^{\mu}$ with $1+N>0$ which is equivalent to pure gauge for the skeleton-gauge vector tetrad structure on plane two. The group $PGB1$ is of measure zero and it is just gradients of scalars in a local plane in a four-dimensional spacetime. There is an isomorphism between the group of local electromagnetic gauge transformations minus the set $PGB1$ and LB2.

\item In reference \cite{ROMP} it was found that the map between $U(1)$ and $LB1 \otimes LB2$ is an isomorphism. The Kernel of this map will be just constant gauge transformations.

\end{itemize}

The mapping between $U(1)$ and $LB1 \otimes LB2$ will be an isomorphism. In the general sense this isomorphism is piecewise. Because we have in the local plane one boosts which are hyperbolic rotations, boosts composed with full inversions which are the composition of two reflections, boosts composed with spacetime reflections and boosts composed with full inversions and spacetime reflections. In the local plane two we have spatial rotations. It is in a general sense a piecewise isomorphism, see reference \cite{ROMP}.


\subsection{PGB1 isomorphic to PGB2}
\label{pgiso}

We would like to know if we can establish a one to one relationship between the groups $PGB1$ and $PGB2$. Let us consider the following approach to this proof. We will use this result later on.

\subsubsection{Coordinate expansions}
\label{coordexp}

Let us suppose solely for the purpose of this section that we establish as coordinates for local scalars $\Lambda$ that only have non-zero gradients in the local plane one $x^{0},x^{1}$. Similar for the plane two for $x^{2},x^{3}$. That is, the objects in $PGB1=\{\Lambda / \Lambda_{\mu} \in \mbox{local Plane 1} \}$ have as only non-zero components of gradients of scalars, $\partial_{0}\Lambda$ and $\partial_{1}\Lambda$. The objects in $PGB2=\{\Lambda / \Lambda_{\mu} \in \mbox{local Plane 2} \}$ have as only non-zero components of gradients of scalars, $\partial_{2}\Lambda$ and $\partial_{3}\Lambda$. Then, we can write in general,

\begin{eqnarray}
\Lambda_{plane1} &=& \sum_{n=0}^{\infty}\sum_{m=0}^{\infty}\:a_{nm}\:(x^{0})^{n}\:(x^{1})^{m} \label{exp1}\\
\Lambda_{plane2} &=& \sum_{n=0}^{\infty}\sum_{m=0}^{\infty}\:b_{nm}\:(x^{2})^{n}\:(x^{3})^{m} \ . \label{exp2}
\end{eqnarray}

We notice that we have the availability of any sequence $a_{nm}$ and a similar availability for any sequence $b_{nm}$ as long as both local scalars $\Lambda_{plane1}$ and $\Lambda_{plane2}$ belong to the local group of electromagnetic transformations. Therefore for any sequence $a_{nm}$ there will be an equal sequence $b_{nm}=a_{nm}$ in the set of all possible sequences for (\ref{exp1}-\ref{exp2}) modulo possible dimensional constants. We can establish very clearly a one to one mapping between $PGB1$ and $PGB2$. We will establish the following conventions that we will use in the remainder of this manuscript:

\begin{enumerate}

\item We will address the isomorphism between the group of local electromagnetic transformations minus the local group $PGB2$ and LB1 as just the isomorphism between the group of local electromagnetic transformations and LB1.

\item We will address the isomorphism between the group of local electromagnetic transformations minus the local group $PGB1$ and LB2 as just the isomorphism between the group of local electromagnetic transformations and LB2.

\item The objects in $PGB2$ map into the kernel of LB1. The objects in $PGB1$ map into the kernel of LB2. The mapping between the elements in the local group $PGB2$ and the local plane two is not trivial. The mapping between the elements in the local group $PGB1$ and the local plane one is not trivial either. Since we have established a one to one mapping between the local groups $PGB1$ and $PGB2$ then we are establishing a mapping between the sector in LB1 that corresponds to the kernel of the mapping into LB2 and the sector in LB2 that corresponds to the kernel of the mapping into LB1. When an element $\Lambda$ is mapped into the kernel of the map into the local plane two, it is also mapped into a possibly non-trivial element of LB1 and vice-versa. We have found a map that takes an element in LB1 that is not trivial (and corresponds to an element in the kernel of LB2) into an element in LB2 that is not trivial (and corresponds to an element in the kernel of LB1) through the isomorphic map between $PGB1$ and $PGB2$. Therefore when we say in the next section \ref{switchinvolution} ``LB1 proper connected to the identity plus the point at infinity is isomorphic to SO(2)=LB2 as groups'' it is understood that we have mapped isomorphically the image of $PGB1$ in the local plane one, that is, into LB1 into the image of $PGB2$ in the local plane two, that is, into LB2. We know that the elements in the local group of electromagnetic gauge transformations with non-trivial image in LB1 and LB2 map an element in LB1 into the corresponding element in LB2, and that will be a natural mapping. We will not repeat these clarifications and these results will remain understood, for the purpose of simplicity and non-redundancy of explanations.

\end{enumerate}

\newtheorem {guesslb2} {Theorem}
\begin{guesslb2}
The local groups $PGB1$ and $PGB2$ are isomorphic.
\end{guesslb2}

\section{The switch is a group involution in LB1}
\label{switchinvolution}

The point has been raised that it is inappropriate to call improper the reflection transformations composed with boosts or even with full inversions since improper Lorentz transformations preserve the timelike or spacelike character of the vectors, while the swap $(0 1|1 0)$ and the composition of the swap reflection with any boost does not preserve it. So the swap is not even improper. Since according to this correct argument we would have to say a boost composed with a reflection or a boost composed with a full inversion and a reflection every time we talk about these kinds of transformations we propose to shorten this denomination by calling them special improper, just for short and encompassing. In manuscript \cite{A} the special improper transformations for the local group of Abelian electromagnetic gauge transformations that mapped into the local plane one were established in equations (62-63), (64-65) and (66-67). They were further analyzed in an Erratum \cite{A}. However, we consider that we can be more precise and complete in the analysis of these mappings. The equations for the special improper case (62-63), (64-65) and (66-67) are correct as are equations (58-59) and (60-61) for the proper case. This analysis is not about the correctness of these equations, it is about their interpretation. For equations (64-65) in reference \cite{A} it was written that ``For $D > 0$ and $1+C > 0$ these transformations (64-65) represent improper space inversions on blade one. If $D > 0$ and $1+C < 0$, equations (64-65) are improper time reversal transformations on blade one \cite{WE}''. For equations (66-67) it was written ``For $D < 0$ and $1+C < 0$ these transformations (66-67) represent the composition of inversions, and improper space inversions on blade one. If $D < 0$ and $1+C > 0$, equations (66-67) are inversions composed with improper time reversal transformations on blade one \cite{WE}''. As it was expressed in the Erratum \cite{A} all these statements are inaccurate. They are also incomplete. These local tetrad vector transformations are special improper because their determinant is $-1$ and at the same time they are not Lorentz transformations because they fail to satisfy the Lorentz equations given by $\Lambda^{\gamma}_{\:\:\alpha}\:\eta_{\gamma\delta}\:\Lambda^{\delta}_{\:\:\beta} \neq \eta_{\alpha\beta}$. For example, in manuscript \cite{A} the special improper case $-1 \leq {(1+C) \over \sqrt{-(1+C)^{2} + D^{2}}} \leq 1$ was never analyzed and it is nonetheless a possible case. Let us do the analysis now to see what we get. As we are dealing with an special improper case we assume the particular situation where $D > (1+C)  > 0$. Therefore, we will discuss $0 < {(1+C) \over \sqrt{-(1+C)^{2} + D^{2}}} < 1$. By performing simple algebraic work on this last inequality we arrive at the condition $0 < \sqrt{2}\:(1+C) < D$ which is not in contradiction with our original assumption $D > (1+C)  > 0$. Now, let us suppose that $(1+C)  < 0$ and $|D| > |1+C|$. We are now considering the case $-1 < {(1+C) \over \sqrt{-(1+C)^{2} + D^{2}}} < 0$. By performing algebraic manipulations on this last inequality we would arrive at $|D| > \sqrt{2}\:|1+C|$ which is not incompatible with our initial assumption $|D| > |1+C|$. These are all possible cases of the special improper nature because of the $-1$ determinant and at the same time they are not Lorentz transformations as stated above. We summarized all these cases in the Erratum \cite{A} by saying that we can represent the transformations (62-63) and (64-65) given in paper \cite{A} as the composition of boosts and the non-Lorentzian discrete transformation that we called the switch. The switch or flip is given by $\Lambda^{o}_{\:\:o} = 0$, $\Lambda^{o}_{\:\:1} = 1$, $\Lambda^{1}_{\:\:o} = 1$,  $\Lambda^{1}_{\:\:1} = 0$ and it is just a reflection. A reflection is not a Lorentz transformation. We made later on, the statement that the group LB1 composed by the boosts plus two discrete transformations, minus the identity two by two, and the switch, is isomorphic to the group SO(2). This last statement would imply that a group like LB1 with three involutions is isomorphic to SO(2)=LB2 with only two involutions. This statement is evidently incomplete and misleading. We will advance the notion that there are two sheets and four subsheets in LB1. One subsheet is connected to the identity. Let us call the first sheet LB1 proper. LB1 proper is made up of the boosts which are connected to the identity plus the boosts composed with minus the identity which are not connected to the identity but which are made with proper Lorentz transformations. These are two subsheets.

We will prove that LB1 proper connected to the identity plus the point at infinity is isomorphic to SO(2)=LB2 as groups. Both have two involutions, which happen to be the same and we will prove in the next section \ref{fullinv} that they correspond through the mapping. Minus the identity in LB1 proper will be mapped into minus the identity in SO(2). We will prove that LB1 proper disconnected to the identity plus the point at infinity is isomorphic to SO(2)=LB2 as groups as well. Then we have the second sheet with two subsheets that we will call LB1 special improper. LB1 special improper is made up by the switch or flip composed with LB1 proper. It is obviously not a group since it does not have the identity, and it is not Lorentzian since the flip is not a Lorentz transformation. LB1 special improper is isomorphic to LB1 proper as sets, not groups. For every boost or boost composed with minus the identity there is a corresponding transformation in LB1 special improper by multiplying by the switch or flip. It is evidently an isomorphism between sets but not groups. Evidently as well there will be the isomorphism between LB1 proper connected to the identity plus the point at infinity multiplied by the reflection and SO(2). As sets, not as groups. We observe similarities with the homomorphism between $SU(2)$ and $SO(3)$. The difference is that here we have two sheets and every sheet has two subsheets. There is a four covering of $SO(2)$. It will be clear through the analysis in the next section that we need to add the point at infinity to LB1 proper connected to the identity in order to establish a group isomorphism with SO(2) and similar for the other three subsheets. The point at infinity for the connected component happens to originate from the unique inhomogeneous map into the future null cone as found in sections \ref{diffeq}-\ref{diffeqrn}-\ref{geneq}.

\section{Homomorphic Map}
\label{mapphomomorph}

It is already understood from section \ref{switchinvolution} that LB1 proper is isomorphic to LB1 special improper, simply because LB1 special improper is found by multiplying or compose LB1 proper with the switch or flip. For the sake of having more insight let us study now the converse problem. Let us give a $\tanh(\phi)$ for some given local scalar $\phi$ and see how the differential equations producing the set LB1 proper relate to the differential equations producing the set LB1 special improper. We know from reference \cite{A} that for LB1 proper we have the coefficient relationship $D = (1+C)\:\tanh(\phi)$ while for LB1 special improper the coefficient relationship becomes $D = (1+C)\:/\:\tanh(\phi)$. Once the tetrad vectors are given and the scalars $\tanh(\phi)$ are also given these are both differential equations in the mapping local scalars $\Lambda$ associated to electromagnetic local gauge transformations. Let us write the differential equations explicitly for both sheets using the $C$ and $D$ coefficients from equations (\ref{COEFFCAT1RN}-\ref{COEFFDAT2RN}),

\begin{eqnarray}
(-Q/2)\:V_{(2)\sigma}\:\Lambda^{\sigma} / (\:V_{(1)\beta}\:
V_{(1)}^{\beta}\:) &=& [1 + (-Q/2)\:V_{(1)\sigma}\:\Lambda^{\sigma} / (\:V_{(2)\beta}\:
V_{(2)}^{\beta}\:)]\:\tanh(\phi)  \label{DIFFEQPROPER} \\
(-Q/2)\:V_{(2)\sigma}\:\Lambda^{\sigma} / (\:V_{(1)\beta}\:
V_{(1)}^{\beta}\:) &=& [1 + (-Q/2)\:V_{(1)\sigma}\:\Lambda^{\sigma} / (\:V_{(2)\beta}\:
V_{(2)}^{\beta}\:)]\:/\tanh(\phi) \label{DIFFEQIMPROPER}
\end{eqnarray}

Equation (\ref{DIFFEQPROPER}) for the proper case and equation (\ref{DIFFEQIMPROPER}) for the special improper case. Given the tetrad and given the local scalar $\phi$, then for the inhomogeneous case there will be one solution $\Lambda_{proper}^{inh}$. Similar for the inhomogeneous $\Lambda_{special\:improper}^{inh}$. Therefore, through the local scalars $\tanh(\phi)$ we can establish a one to one correspondence between the proper and the special improper families of solutions. A one to one relation between the proper and the special improper sheets. We can also manage to rewrite the homogeneous version of equations (\ref{DIFFEQPROPER}-\ref{DIFFEQIMPROPER}) as,

\begin{eqnarray}
V_{(2)\sigma}\:\Lambda_{proper}^{h\sigma} + V_{(1)\sigma}\:\Lambda_{proper}^{h\sigma}\:\tanh(\phi) &=& 0  \label{HOMODIFFEQPROPER} \\
V_{(2)\sigma}\:\Lambda_{special\:improper}^{h\sigma} + V_{(1)\sigma}\:\Lambda_{special\:improper}^{h\sigma}\: {1 \over \tanh(\phi)} &=& 0 \ , \label{HOMODIFFEQIMPROPER}
\end{eqnarray}

where the upstairs h stands for homogeneous. We can rewrite equations (\ref{HOMODIFFEQPROPER}-\ref{HOMODIFFEQIMPROPER}) as,

\begin{eqnarray}
\tanh(\phi) &=& - {V_{(2)\sigma}\:\Lambda_{proper}^{h\sigma} \over V_{(1)\tau}\:\Lambda_{proper}^{h\tau}}  \label{HOMODIFFEQPROPER2} \\
\tanh(\phi) &=& - {V_{(1)\sigma}\:\Lambda_{special\:improper}^{h\sigma} \over V_{(2)\tau}\:\Lambda_{special\:improper}^{h\tau}} \ , \label{HOMODIFFEQIMPROPER2}
\end{eqnarray}

We can think numerators and denominators in both equations (\ref{HOMODIFFEQPROPER2}-\ref{HOMODIFFEQIMPROPER2}) as the orthogonal sides in the right triangles associated to hyperbolas. It is simple to see that these ratios will be the same modulo constant factors like $a\:\Lambda_{proper}^{h\sigma}$ or $b\:\Lambda_{special\:improper}^{h\sigma}$ since they are just factors of similarity in similar right triangles. We must emphasize that since we found that the Kernel for the proper case, for example associated to equation (\ref{HOMODIFFEQPROPER2}) is just $PGB2$ isomorphic to $PGB1$, then if the choice for gauge vector $X^{\mu}$ in the tetrad vectors that generate the local plane one (\ref{V1}-\ref{V2}) is not pure gauge, the numerators and denominators in equation (\ref{HOMODIFFEQPROPER2}) will not be trivial as well. In the local plane two we would have an absolutely similar analysis for $\tan(\varphi) = {V_{(3)\sigma}\:\Lambda^{h\sigma} \over V_{(4)\tau}\:\Lambda^{h\tau}}$.

\subsection{Several important points}
\label{severalpoints}

Let us make some checks on the equations that define the local tetrad transformations for the special improper case, starting with the case $D_{o} > 1 + C_{o}$, $0 > C_{o} > -1$. We can always have this case as it is proven in the section VIII in reference \cite{ASU3}. We can consider for the special improper equations (64-65) in reference \cite{A} a new case where $0 < 1 + C_{new} = \epsilon_{n} \ll 1$, $n$ natural, for example we can have a new $\Lambda_{new} = A_{n} \: \Lambda$ for some positive $A_{n}$ where $\Lambda$ is the old local scalar. Then $1 - A_{n}\:|C_{o}| = 1 + C_{new} = \epsilon_{n}$ where $D_{new} = A_{n}\:D_{o} > 1 + C_{new} = 1 -A_{n}\: |C_{o}| > 0$. If we can make $\epsilon_{n} \rightarrow 0$ as $n \rightarrow \infty$, then this is equivalent to making the limit in equations (64-65) in reference \cite{A} equal to the switch or flip. Therefore, the switch or flip is an actual element in LB1. We can see this last result under a new light by considering the corresponding differential equation having a particular inhomogeneous local electromagnetic gauge scalar solution $\Lambda_{special\:improper}^{inh}$ as,

\begin{eqnarray}
[1 + C_{special\:improper}^{inh}] = [1 + (-Q/2)\:V_{(1)\sigma}\:\Lambda_{special\:improper}^{inh\:\:\sigma} / (\:V_{(2)\beta}\:V_{(2)}^{\beta}\:)] = 0 \label{DIFFEQIMPROPERFLIP}
\end{eqnarray}

The differential equation (\ref{DIFFEQIMPROPERFLIP}) has only one inhomogeneous $\Lambda_{special\:improper}^{inh}$ solution which corresponds to the flip or switch. Therefore the coefficient $D_{special\:improper}^{inh}$ in equations (64-65) in reference \cite{A} has to be positive. It has to be positive because the second sheet is just the flip composed with the first sheet and since the identity truly belongs to the first sheet, the differential equation (\ref{DIFFEQIMPROPERFLIP}) has only one $\Lambda_{special\:improper}^{inh}$ solution which corresponds to the $D_{special\:improper}^{inh} > 0$ case.
We will show next that we can even construct a sequence converging towards minus the identity inside LB1 and similar in LB2 and it will be proved in section \ref{fullinv} that it is a simultaneous limit accumulation point in both LB1 and LB2. We will prove that minus the identity in both LB1 and LB2 is a simultaneous accumulation point in both groups. Therefore, similarly for equations (66-67) in reference \cite{A} we can proceed as follows. We can consider a proper case $1 + C_{o} > D_{o}$ for $C_{o} > 0$ and like the previous case consider a new transformation $\Lambda_{new} = -A \: \Lambda$ such that $D_{new} = -A\:D_{o} < 0$ and $1 + C_{new} = 1 - A\:C_{o} < 0$ where now $C_{o} > D_{o} > 0$ and $A$ is positive. Then the case will be special improper with $|D_{new}| = |-A\:D_{o}| > |1 -A\: C_{o}| > 0$ making $A < 1 / (C_{o}-D_{o})$. We can make $-1 + A_{n}\: C_{o} = \epsilon_{n}$ for $\epsilon_{n} \ll 1$ with $n$ natural. When we take the limit of a sequence $\epsilon_{n} \rightarrow 0$ as $n \rightarrow \infty$, then this is equivalent to making the limit in equations (66-67) in reference \cite{A} equal to minus the switch or flip. Let us suppose now that we choose non-trivial gauge vectors $X^{\mu}$ and $Y^{\mu}$ in the tetrad construction (\ref{V1}-\ref{V4}). Then, as we saw in section \ref{kernelmap} the elements of the local group of Abelian scalars $\Lambda$ that will map into the identity are the constants and the zero measure subgroups $PGB1$ isomorphic to $PGB2$ as long as the gauge vectors are not scalars multiplied by pure gauge. On both local planes or blades one and two. Then, the minus identity in LB1 proper cannot map into the identity in SO(2). Let us remember that the mappings between the local group of electromagnetic gauge transformations and the local LB1 and separately LB2 local groups of tetrad transformations have been proved surjective, see references \cite{A,ASU3}. Surjective in the sense that the image of the mapping into the local plane one is not a subgroup of the codomain LB1. Moreover, since the identity is mapped only by constants in both LB1 and LB2=SO(2), then the only involution left for minus the identity in LB1 proper to be mapped into SO(2) is minus the identity two by two in SO(2). Therefore, the identity in LB1 proper corresponds to the identity in LB2=SO(2) and minus the identity in LB1 proper corresponds to minus the identity in SO(2). We will prove the correspondence for the full inversion or minus the identity in section \ref{fullinv}. LB1 proper connected to the identity plus the point at infinity is mapped by transitivity into SO(2). The issue of the point at infinity will be shown with transparency once we employ stereographic projections in section \ref{infinity}. Since we proved by several means that LB1 special improper is as a set isomorphic to LB1 proper, then as a set LB1 special improper plus the point at infinity will be isomorphic to SO(2) as well. Since the identity in LB1 proper is mapped into the switch or flip in LB1 special improper, then the switch or flip is mapped into the identity in SO(2). Since minus the identity in LB1 proper is mapped into minus the flip in LB1 special improper, then, minus the flip is mapped into minus the identity in SO(2). The surjectivity or onto nature of the map between LB1 plus the point at infinity and LB2=SO(2) was well proved in manuscript \cite{A}, and we must say that in manuscript \cite{ASU3} we further exemplified in Appendix III this surjectivity as it was already explained above. Therefore we conclude this section by stating that there is a homomorphism between LB1 plus the four points at infinity and LB2=SO(2). LB1 proper connected to the identity plus the point at infinity is isomorphic as a group to SO(2). LB1 proper disconnected to the identity plus the point at infinity is isomorphic as a set to SO(2). LB1 special improper is isomorphic to LB1 proper. LB1 proper has two involutions as SO(2) has and they map to each other. When we add the points at infinity for both possible solutions on the future light cone and on the past light cone we obtain as we will see in section \ref{infinity} a two to one homomorphism between LB1 proper and LB2. LB1 special improper has also two involutions and they map to the two involutions of SO(2) one to one. When we add the points at infinity for both possible solutions on the future light cone and on the past light cone for the reflected null solutions which are the same as for the unreflected case we obtain as we will see in section \ref{infinity} a two to one homomorphism between LB1 special improper and LB2. There is for the Kernel of the mapping between LB1 and SO(2) two elements $\{identity, switch\}$ if the gauge vectors are not chosen to be pure gauge and we exclude the zero measure subgroups $PGB2$ isomorphic to $PGB1$, see reference \cite{ROMP}. Let us remember that in section \ref{kernelmap} we only studied the Kernel of LB1 proper which was trivial of measure zero. Both discrete transformations in LB1 $\{identity, switch\}$ are mapped into the identity in SO(2). This is clearly a homomorphism and a double covering of SO(2) by LB1 proper. We summarize the results in the past two sections,

\begin{guesslb2}
The mapping between the local group of tetrad transformations on the local blade one LB1 corresponding to equations (58-59) (60-61) (64-65) (66-67) in manuscript \cite{A} plus the four points at infinity and the local group of tetrad transformations on the local orthogonal blade two LB2=SO(2) given by equations (91-92) in manuscript \cite{A} is a homomorphism. LB1 plus the four points at infinity is a four covering of SO(2) and the Kernel of this mapping is given by $\{identity, switch=flip\}$. The LB1 subgroup LB1 proper connected to the identity plus the point at infinity is isomorphic to SO(2). The points at infinity correspond to the four inhomogeneous gauge solutions found in sections \ref{diffeq}-\ref{diffeqrn}-\ref{geneq}.
\end{guesslb2}

The issue of the point at infinity will be shown with transparency once we employ stereographic projections in section \ref{infinity}.


\section{The full inversion is a simultaneous accumulation point in LB1 and LB2}
\label{fullinv}

We would like to add that we will investigate in detail in a forthcoming paper the nature of the group law which is not trivial for this mapping between the local group of electromagnetic gauge transformations and the local groups LB1 and LB2. However we can advance that the group law is satisfied for the mapping under study. The analysis cannot be included in this manuscript because it is too long when we exhaust all possible cases even though in section \ref{appendixII} we apply the group law to the composition of two boosts or boosts composed with full inversions.
However, we would like to study the question about why we say that the full inversion is a group element in LB1 and independently in LB2 connected through the mapping, mapping to each other transitively. Let us investigate this issue in more detail because there is a reason. The proof will consist of a two stage limit process. Let us start by assuming that we consider a local boost with $1 + C_{o} > D_{o}$, $C_{o} > D_{o} > 0$. This is a LB1 proper local transformation of the kind (58-59) in reference \cite{A}, they exist and it is valid to consider such a case. Then, it is also valid to consider the new case where the generating local electromagnetic scalar $\Lambda$ is changed into $-n\:\Lambda$ where $n$ is a natural number. The transformation coefficients are changed into $-n\:C_{o}$ and $-n\:D_{o}$. We would like the new case to be a proper Lorentz transformation on the local plane one of the kind (60-61) in reference \cite{A}. Therefore we will have $|1-n\:C_{o}| > |-n\:D_{o}|$, $1-n\:C_{o} < 0$ for n sufficiently large and $n \rightarrow \infty$. We have turned a (58-59) boost case into a (60-61) case, a boost composed with minus the identity. We then rewrite the coefficient of the Lorentz transformation on plane one as,

\begin{eqnarray}
{(1+C) \over \sqrt{(1+C)^2-D^2}} = {(1-n\:C_{o}) \over \sqrt{(1-n\:C_{o})^2-(-n\:D_{o})^2}} = {-1 \over \sqrt{1-({-n\:D_{o} \over 1-n\:C_{o}})^2}} \ . \label{newc}
\end{eqnarray}

For the natural $n$ sufficiently large we would be able to approximate equation (\ref{newc}) to,

\begin{eqnarray}
{-1 \over \sqrt{1-({-n\:D_{o} \over 1-n\:C_{o}})^2}} \rightarrow {-1 \over \sqrt{1-({D_{o} \over C_{o}})^2}} \ . \label{newcc}
\end{eqnarray}

We next consider for the second stage limit process the full expression for ${D_{o} \over C_{o}}$ given in equation (\ref{HOMODIFFEQPROPER2}),

\begin{eqnarray}
{D_{o} \over C_{o}} = - {V_{(2)\sigma}\:\Lambda_{proper-plane1}^{h\sigma} \over V_{(1)\tau}\:\Lambda_{proper-plane1}^{h\tau}} = \tanh(\phi) \label{finalaccum1}
\end{eqnarray}

for some $\Lambda_{proper-plane1}^{h\sigma}$ and some $\tanh(\phi)$. As a comment, we can repeat in parallel this whole analysis simultaneously for the equations (91-92) in reference \cite{A} and the coefficient ${(1+N) \over \sqrt{(1+N)^2+M^2}}$ on the local blade or plane two and get through equation $\tan(\varphi) = {V_{(3)\sigma}\:\Lambda^{h\sigma} \over V_{(4)\tau}\:\Lambda^{h\tau}}$ an analogous result for the local plane two,

\begin{eqnarray}
{M_{o} \over N_{o}} = {V_{(3)\sigma}\:\Lambda_{proper-plane2}^{h\sigma} \over V_{(4)\tau}\:\Lambda_{proper-plane2}^{h\tau}} = \tan(\varphi) , \label{finalaccum2}
\end{eqnarray}

for some $\Lambda_{proper-plane2}^{h\sigma}$ and some $\tan(\varphi)$. Since all these results occur when we choose a non-pure-gauge gauge vector, the numerators and denominators in equations (\ref{finalaccum1}-\ref{finalaccum2}) will be non-trivial. Even more so if we also exclude the zero measure cases $PGB2$ and $PGB1$. It is relevant to mention that once we took the first limit when we did ${-1 \over \sqrt{1-({-n\:D_{o} \over 1-n\:C_{o}})^2}} \rightarrow {-1 \over \sqrt{1-({D_{o} \over C_{o}})^2}},\: for \: n \rightarrow \infty$, we moved from the inhomogeneous equation (\ref{DIFFEQPROPER}) into the homogeneous case (\ref{HOMODIFFEQPROPER2}). Observe that ${-n\:D_{o} \over 1-n\:C_{o}} \rightarrow {D_{o} \over C_{o}},\: for \: n \rightarrow \infty$. Similar when we dealt with the analogous problem on the orthogonal plane two for the inhomogeneous equation

\begin{eqnarray}
(-Q/2)\:V_{(3)\sigma}\:\Lambda^{\sigma} / (\:V_{(4)\beta}\:V_{(4)}^{\beta}\:) = [1 + (-Q/2)\:V_{(4)\sigma}\:\Lambda^{\sigma} / (\:V_{(3)\beta}\:
V_{(3)}^{\beta}\:)]\:\tan(\varphi) \ , \label{HOMODIFFERPLANE2}
\end{eqnarray}

into the homogeneous case $V_{(3)\sigma}\:\Lambda_{space}^{h\sigma} - V_{(4)\sigma}\:\Lambda_{space}^{h\sigma}\:\tan(\varphi) = 0$. It must be emphasized that for the other local boost matrix component of the LB1 transformation considered at the outset of this section, ${D \over \sqrt{-(1+C)^2+D^2}}$ from equations (58-59) in reference \cite{A}, the final limit after the two stage limit process would be zero. Analogous for the other matrix component of the local LB2 spatial rotation, ${M \over \sqrt{(1+N)^2+M^2}}$ from equations (91-92) in reference \cite{A} in the local orthogonal plane two. Then, the next step in our proof involves the fact that we can always consider a sequence or succession of local scalars $\phi_{m}$ and $\varphi_{m}$ for $m$ a natural number such that save for possible multiplicative constants $c$ we can make the choice $\phi_{m} = c\:\varphi_{m}$. Multiplicative constants $c$ could be important since $\phi$ might be adimensional while $\varphi$ might be given in non-adimensional units, for example. This sequence will be chosen to tend to local scalar fields $\phi_{m} = c\:\varphi_{m} \rightarrow 0$ when $m \rightarrow \infty$. This is possible since $\tanh(0) = \tan(0) = 0$. It is clear that for each $m$ and each given $\phi_{m}$ and $\varphi_{m}$ equations (\ref{finalaccum1}-\ref{finalaccum2}) will become differential equations separately and independently in $\Lambda_{m\:proper-plane1}^{h\sigma}$ and $\Lambda_{m\:proper-plane2}^{h\sigma}$ for every $m$. In the set of local tetrad group elements both in LB1 and LB2 there will be for every open ball with center in the full inversion an infinite set or sequence of group elements as we have proved in this section. We have that the sequence in both local orthogonal planes one and two was chosen as to have common scalars in $\phi_{m} = c\:\varphi_{m} \rightarrow 0$ when $m \rightarrow \infty$. Therefore, we must conclude that when we follow both sequences in both local orthogonal planes associated to both LB1 and LB2 local groups of tetrad transformations, we have the full inversion in both groups LB1 and LB2 as an accumulation point. Every open ball centered in the full inversion contains an infinite number of group elements in both LB1 and LB2, therefore the full inversion in both local groups of tetrad transformations LB1 and LB2 is a limit point and an accumulation point. Since in both local groups LB1 and LB2 we can construct a one to one sequence converging into the full inversion. Because it is a common limit point of a one to one sequence from LB1 into LB2 such that it is a simultaneous accumulation point for both groups.
Therefore, the full inversion not only belongs to the mapping image in LB1 or LB2 but it is also an accumulation point in both local groups, simultaneously. We are proving simultaneously along with the analysis in section \ref{severalpoints} that minus the flip is an accumulation point in LB1 in a one to one sequence with the sequences just presented.

\section{Infinity limits}
\label{infinity}

Let us consider the plane with rectangular coordinates $(X,Y)$ and on this plane we will also consider the hyperbola $X^{2}-Y^{2}=1$. But for simplicity we will only consider the upper half branch $X\geq0$ and $Y\geq0$. We intend to study the possible existence of a map between this upper half branch and the quadrant $X\geq0$ and $Y\geq0$ for the circle $X^{2}+Y^{2}=1$. The method that we will use involves the unit sphere $S^{2}$ and the stereographic projection through the north pole. We will study the hyperbola and the circle on a plane that cuts the sphere through the equator in such a way that the circle is the equator of the unit sphere $S^{2}$. Let us introduce the coordinates of the stereographic projections for the unit sphere $S^{2}$. The local 2-sphere is defined through $\sum_{i=1}^{3} x_{i}^{2} = 1$ where the $x_{i},\: i=1 \cdots 3$ are local coordinates. Following closely chapter III in \cite{CBDW} an in order to construct an atlas we let $P$ and $Q$ be the north and south poles respectively. Let $U=S^{2}-{P}$ and $V=S^{2}-{Q}$, let g and h be the stereographic projections of the poles $P$ and $Q$ on the plane $x_{3}=0$,

\begin{center}
$\: g: U \rightarrow \Re^{2} \:\:\: by  \:\:\:  y_{i} = x_{i} / (1-x_{3}) \:\:\: for  \:\:\:  i=1 \cdots 2$
\end{center}

\begin{center}
$\: h: V \rightarrow \Re^{2} \:\:\: by  \:\:\: z_{i} = x_{i} / (1+x_{3}) \:\:\: for  \:\:\: i=1 \cdots 2$
\end{center}

See reference \cite{CBDW} for the proof that this is an atlas. Let us parameterize the upper half branch of the hyperbola under consideration by $(\cosh(t),\sinh(t))$ with $t:0 \rightarrow +\infty$. In turn we will parameterize the unit circle on the same plane quadrant with $(\cos(\varphi),\sin(\varphi))$ and with $\varphi:0 \rightarrow \pi/2$. Then, if the coordinates on the sphere are $(x,y,z)$ and the coordinates on the plane are $(X,Y)$ the relationship between them through the stereographic projection of the north pole will be,

\begin{center}
$(X,Y) = ({x \over 1-z},{ y \over 1-z})$
\end{center}

\begin{center}
$(x,y,z) = ({2\:X \over 1+X^{2}+Y^{2}},{ 2\:Y \over 1+X^{2}+Y^{2}}, {-1+X^{2}+Y^{2} \over 1+X^{2}+Y^{2}})$
\end{center}

If in the second of these two stereographic coordinate relationships we set $(X=\cosh(t),Y=\sinh(t))$ we obtain,

\begin{center}
$(x,y,z) = ({2\:\cosh(t) \over 1+\cosh^{2}(t)+\sinh^{2}(t)},{ 2\:\sinh(t) \over 1+\cosh^{2}(t)+\sinh^{2}(t)}, {-1+\cosh^{2}(t)+\sinh^{2}(t) \over 1+\cosh^{2}(t)+\sinh^{2}(t)})$
\end{center}

We might also notice that $\cosh^{2}(t)+\sinh^{2}(t)=\cosh(2t)$. Finally we establish the following mapping,

\begin{eqnarray}
\cos(\varphi) &=& {2\:\cosh(t) \over 1+\cosh^{2}(t)+\sinh^{2}(t)} = {2\:\cosh(t) \over 1+\cosh(2t)} \label{stereomap}
\end{eqnarray}

Within the range $t:0 \rightarrow +\infty$ and $\varphi:0 \rightarrow (\pi/2)^{-}$ this map is truly and isomorphism. We can immediately notice that the infinity value for the coordinate t will correspond to the north pole and the isomorphic map will not reach $\pi/2$ even though $\pi/2$ is an accumulation point. Then the issue of the point at infinity arises. It is evident that it must be added in order to reach the point $\cos\varphi=\pi/2$. This map is telling us that the upper branch for $X\geq0$ and $Y\geq0$ maps into the first quadrant of the circle except for the point $\varphi=\pi/2$ which is an accumulation point. It happens that the point at infinity is the point in the asymptote corresponding to the future light cone found in sections \ref{diffeq}-\ref{diffeqrn}-\ref{geneq}. There is a unique inhomogeneous solution as in sections \ref{diffeq}-\ref{diffeqrn}-\ref{geneq} such that a unique local electromagnetic gauge transformation is mapped into tetrad vectors lying on the local future light cone. Since the ``point at infinity'' is the point on the asymptote of the upper branch of the hyperbola $X^{2}-Y^{2}=1$, then this point exists in the image of the map that we are studying. This is the point that maps into the north pole of the stereographic unit sphere. It is the electromagnetic gauge transformation that solved the inhomogeneous equation $D=1+C$ in sections \ref{diffeq}-\ref{diffeqrn}-\ref{geneq}. Analogous for the lower branch $X\geq0$ and $Y\leq0$ and also analogous for the opposite branch with $X\leq0$ and $Y\leq0$ and also $X\leq0$ and $Y\geq0$ with the unique solution for $D=-(1+C)$. The inhomogeneous solution for $D=-(1+C)$ is in the past light cone and corresponds to the point at infinity for the opposite branch of the hyperbola. Once we consider the point at infinity plus the upper and lower branches for both opposite branches of the hyperbola we close the curve at the north pole and we would have two closed curves. It is simple to see that we can also repeat the whole argument for the conjugate hyperbola $Y^{2}-X^{2}=1$ and its two branches. The conjugate hyperbola is a reflection of the original hyperbola through the asymptote $Y=X$. The reflected inhomogeneous solutions for the past and future light cones are the same as for the original hyperbola since they are invariant by reflection through the asymptote $Y=X$. By topological closure we would be able to map the conjugate hyperbola plus two infinities corresponding to $D=1+C$ and $D=-(1+C)$ we will have two closed curves mapped into $SO(2)$. A total of four closed curves mapped into $SO(2)$ when also considering the conjugate hyperbola. The group LB1 is given by $SO(1,1) \times Z_{2} \times Z_{2}$ where $SO(1,1)$ is proper orthochronous. The first $Z_{2}$ is given by $\{I_{2 \times 2}, -I_{2 \times 2}\}$ and the second $Z_{2}$ is given by $\{I_{2 \times 2}, \mbox{the swap}\: (01|10)\}$. We would have to add in order to complete the image of the map $SO(1,1) \times Z_{2} \times Z_{2}\: \bigoplus \: \{light\:cone\:gauge\}$ where the light cone gauge includes the inhomogeneous two solutions to the differential equations in the local future and past light cones established in sections \ref{diffeq}-\ref{diffeqrn}-\ref{geneq} where the reflection through the asymptote $Y=X$ will produce two more identical inhomogeneous solutions. A total of four. We are suppressing the homogeneous solutions to these differential equations for the possible map of local gauge transformations into local tetrad transformations that take timelike and spacelike tetrad vectors on the local plane one into the intersection of the plane with the local light cone. These isomorphisms will be established modulo homogeneous solutions to the differential equations as found in sections \ref{diffeq}-\ref{diffeqrn}-\ref{geneq}. These arguments allow us to make further analysis. Let us consider the equation (\ref{DIFFEQPROPER})

\begin{eqnarray}
(-Q/2)\:V_{(2)\sigma}\:\Lambda^{\sigma} / (\:V_{(1)\beta}\:
V_{(1)}^{\beta}\:) &=& [1 + (-Q/2)\:V_{(1)\sigma}\:\Lambda^{\sigma} / (\:V_{(2)\beta}\:
V_{(2)}^{\beta}\:)]\:\tanh(\phi) \ . \label{DIFFEQPROPERINFLIM}
\end{eqnarray}

If we now consider a sequence $\tanh(\phi_{n})={1 \over (1+\varepsilon_{n})}$ with $\varepsilon_{n} \rightarrow 0^{+}$ for $n \rightarrow \infty$, then we will have by means of a sequence of differential equations of the kind (\ref{DIFFEQPROPERINFLIM}) a sequence of solutions $\Lambda^{hyperbola}_{n}$. Now we consider equation (\ref{HOMODIFFERPLANE2}),

\begin{eqnarray}
(-Q/2)\:V_{(3)\sigma}\:\Lambda^{\sigma} / (\:V_{(4)\beta}\:V_{(4)}^{\beta}\:) = [1 + (-Q/2)\:V_{(4)\sigma}\:\Lambda^{\sigma} / (\:V_{(3)\beta}\:
V_{(3)}^{\beta}\:)]\:\tan(\varphi) \ . \label{HOMODIFFERPLANE2INFLIM}
\end{eqnarray}

If we once more consider a sequence $\tan(\varphi_{n})={1 \over \varepsilon_{n}}$ with $\varepsilon_{n} \rightarrow 0^{+}$ for $n \rightarrow \infty$, then we will have by means of a sequence of differential equations of the kind (\ref{HOMODIFFERPLANE2INFLIM}) a sequence of solutions $\Lambda^{circle}_{n}$. We end this whole argument by saying that finally through the use of the sequence $\varepsilon_{n} \rightarrow 0^{+}$ when $n \rightarrow \infty$ we establish an isomorphism between the sequences $\Lambda^{hyperbola}_{n}$ and $\Lambda^{circle}_{n}$. We managed to translate the mapping between the hyperbola and the circle into a mapping between local scalars associated to tetrad transformations in LB1 and LB2. Please pay attention to the fact that the result in section \ref{fullinv} found in a laborious way becomes trivial using the technique in this section since from equation (\ref{stereomap}) for $t=0$ we obtain $\varphi=0$ corresponding to the identity and a similar result for the opposite branch with $X\leq0$ and $Y\geq0$ corresponding to minus the identity.

\section{Conclusions}
\label{conclusions}

In the general theory of local gauge transformations of tetrad vectors (\ref{V1}-\ref{V2}) on the local plane one determined precisely by these tetrad vectors, we have that after the gauge transformation the norm of the transformed vectors can be expressed as in equations (56-57) in reference \cite{A},

\begin{eqnarray}
\tilde{V}_{(1)}^{\alpha}\:\tilde{V}_{(1)\alpha} &=&
[(1+C)^2-D^2]\:V_{(1)}^{\alpha}\:V_{(1)\alpha}\\
\tilde{V}_{(2)}^{\alpha}\:\tilde{V}_{(2)\alpha} &=&
[(1+C)^2-D^2]\:V_{(2)}^{\alpha}\:V_{(2)\alpha}\ .
\end{eqnarray}

The coefficients C and D are given by equations (54-55) in reference \cite{A},

\begin{eqnarray}
C&=&(-Q/2)\:V_{(1)\sigma}\:\Lambda^{\sigma} / (\:V_{(2)\beta}\:
V_{(2)}^{\beta}\:)\\
D&=&(-Q/2)\:V_{(2)\sigma}\:\Lambda^{\sigma} / (\:V_{(1)\beta}\:
V_{(1)}^{\beta}\:)\ .
\end{eqnarray}

If the coefficient factor $[(1+C)^2-D^2]$ is positive, then the local tetrad vector transformations are Lorentz proper transformations, see the details in reference \cite{A}. If the coefficient factor $[(1+C)^2-D^2]$ is negative, then the local tetrad vector transformations are Lorentz special improper transformations, see the details in reference \cite{A}. The situation arises that the coefficient might be zero in a particular case. Even though in our original paper \cite{A} we treat the general theory of tetrad gauge transformations as addressing the vacuum Einstein-Maxwell spacetimes with no sources to the Maxwell equations, the Coulomb case in flat spacetime shares similarities with the Reissner-Nordstr\"{o}m case, and therefore we set out in this note to analyze this null case for tetrad vector gauge transformations in plane one. Beyond these considerations we found interesting to discuss the Coulomb problem on the local plane one generated by the vectors (\ref{V1}-\ref{V2}) because in the Reissner-Nordstr\"{o}m case which is a solution to the vacuum Einstein-Maxwell equations a similar situation to the one presented in the Coulomb case arises, and the Coulomb case is simpler to study. Not the same because the spacetime has a different metric tensor, but similar. We found by direct analysis in section \ref{nullcase} that the singular gauge is a particular gauge that takes two non-null vectors into the local light cone. We could establish this gauge as a limiting gauge in a succession $A_{t} = e/r$ and $\Lambda_{r}(n) = -(1-1/n)\:e/r$ with $n\rightarrow \infty$. For every natural value of $n$ this gauge transformation would take the timelike vector (\ref{V1}) and the spacelike (\ref{V2}) into a new timelike and spacelike vectors. Only in the limit they both would become null vectors on the local light cone. Therefore, this limit would obviously become an accumulation point in the infinite set of all local gauge transformations. We also proved that it is a set of zero measure or only the single local gauge transformation $\Lambda_{r} = -e/r$ that causes this situation by solving the general differential equation in section \ref{diffeq}. There were homogeneous solutions to the differential equation in section \ref{diffeq}, but just this single inhomogeneous local gauge transformation $\Lambda_{r} = -e/r = -A_{t}$ from the set of gauge transformations. Let us not forget that there is another inhomogeneous singular solution $\Lambda_{r} = e/r = A_{t}$ arising from the case $D = -(1+C)$ and since the analysis is completely analogous, we omit it but know that this second case is associated to the local past light cone while the case $D = (1+C)$ is associated to the local future light cone. But the singular gauges are two and again, a set of measure zero within the infinite set of local gauge transformations. In section \ref{diffeqrn} we did a completely analogous analysis for the Reissner-Nordstr\"{o}m case. As predicted, we found it to be similar but not the same as the Coulomb case in flat Minkowski spacetime. The considerations for the general case in section \ref{geneq} lead once more to similar results. Therefore, these singular cases are unique in the infinite set of local gauge transformations. The four points at infinity in section \ref{infinity} are exactly these two cases plus the two cases analogous for the conjugate hyperbola. As a second problem we set out to find details of the Kernel of the mapping between the set of electromagnetic local gauge transformations and the LB1 proper sheet. Independently also the Kernel of the mapping between the set of electromagnetic local gauge transformations and LB2=SO(2). The Kernel of these maps have been studied in detail in reference \cite{ROMP}. We found that the Kernel is different from the identity if the gauge vectors are trivial pure-gauge $X^{\rho} = \Lambda^{\rho}$ and $Y^{\rho} = \ast \Lambda^{\rho}$. If the gauge vectors are chosen to be non-trivial then the mapping into LB1 proper and SO(2) separately have as Kernel only the constant electromagnetic gauge scalars $\Lambda = constant = \ast \Lambda$ plus the zero measure sets $PGB2$ and $PGB1$ respectively, see reference \cite{ROMP}. We also proved that the zero measure sets $PGB2$ and $PGB1$ are isomorphic between themselves, see section \ref{kernelmap}. In addition in reference \cite{ROMP} it was found that the map between $U(1)$ the group of electromagnetic gauge transformations and the group $LB1 \bigotimes LB2$ is an isomorphism with trivial Kernel. We also analyzed in section \ref{fullinv} the full inversion as a simultaneous accumulation point in both local groups LB1 and LB2 and concluded that there is a one to one sequence converging simultaneously to the full inversion in both LB1 and LB2. It is the limit of both sequences in LB1 and LB2 of group elements in a one to one relation that converge to the full inversion simultaneously as an accumulation point in both groups. We would like to stress that the mapping from the group of local electromagnetic gauge transformations into both the groups of local tetrad transformations LB1 and LB2 on both local orthogonal planes one and two satisfies the group law as it will be proved in detail in a forthcoming paper. In sections \ref{switchinvolution}-\ref{mapphomomorph} we finally found that the statement LB1 plus two discrete transformations is isomorphic to LB2=SO(2) is inaccurate and misleading. It cannot be that LB1 with three involutions $\{identity, -identity, switch\}$ is isomorphic to SO(2) with only two involutions $\{identity, -identity\}$. We therefore found different ways in these sections to prove that LB1 has two sheets including four subsheets. LB1 proper which are the boosts and the boosts composed with minus the identity two by two as the proper sheet and the special improper second sheet made up of the elements of LB1 proper composed with the flip or switch which is nothing but a reflection and therefore a non-Lorentzian transformation with determinant $-1$. LB1 proper has then only two involutions, the identity mapped into the identity in LB2=SO(2) and minus the identity mapped into the minus the identity in LB2=SO(2). LB1 special improper has two involutions, the switch or flip mapped into the identity in LB2=SO(2) and minus the flip mapped into minus the identity in LB2=SO(2). But then in section \ref{infinity} we add for each of the four subsheets a point at infinity which are essentially the unique local gauge transformations mapping into the local future and past light cones and similar for the special improper two subsheets through reflections. This way we proved that $LB1 = \{LB1\:proper, LB1\:special\:improper \}$ is a four covering of LB2=SO(2) when we also invoke topological closure and include the four points at infinity for the four subsheets as we did in detail in section \ref{infinity}. Similar to the case SU(2) and SO(3). The difference here is that there are two sheets in LB1 and two subsheets in both sheets $\{boosts, boosts-composed-with-minus-the-identity\}$ for LB1 proper and  $\{boosts-composed-with-the-flip, boosts-composed-with-minus-the-identity-and-the-flip\}$ for LB1 special improper. The Kernel of the mapping between LB1 and LB2=SO(2) is $\{identity, switch=flip\}$. Even though LB1 proper connected to the identity plus the point at infinity is isomorphic to SO(2) as groups, LB1 is homomorphic to SO(2). This is most certainly a novel result in group theory.

\section{Appendix I}
\label{sec:appI}

We will study in this section how to make a suitable choice for the gauge vector $Y^{\alpha}$ for the Maxwell equations with a source $J^{\mu}$. Let us focus for practical purposes in the Coulomb problem as an example that permits a better visualization of this physical situation. The point is that in geometrodynamics, the Maxwell equations,

\begin{eqnarray}
f^{\mu\nu}_{\:\:\:\:\:;\nu} &=& J^{\mu} \label{L1}\nonumber\\
\ast f^{\mu\nu}_{\:\:\:\:\:;\nu} &=& 0 \ , \label{L2}
\end{eqnarray}

tell us about the existence of one potential $X^{\alpha}=A^{\alpha}$. Then the question arises about gauging the vectors in equations (\ref{V1NONFIXED}-\ref{V4NONFIXED}). The tetrad of eigenvectors to the stress-energy tensor (\ref{TEMDR}) is given by,

\begin{eqnarray}
V_{(1)}^{\alpha} &=& \xi^{\alpha\lambda}\:\xi_{\rho\lambda}\:X^{\rho}
\label{V1NONFIXEDAPP}\\
V_{(2)}^{\alpha} &=& \sqrt{-Q/2} \:\: \xi^{\alpha\lambda} \: X_{\lambda}
\label{V2NONFIXEDAPP}\\
V_{(3)}^{\alpha} &=& \sqrt{-Q/2} \:\: \ast \xi^{\alpha\lambda} \: Y_{\lambda}
\label{V3NONFIXEDAPP}\\
V_{(4)}^{\alpha} &=& \ast \xi^{\alpha\lambda}\: \ast \xi_{\rho\lambda}
\:Y^{\rho}\ .\label{V4NONFIXEDAPP}
\end{eqnarray}

When we make the gauge choice  $f_{tr} = e/r^{2}$, $A_{t} = e/r$ and $A_{r} = 0$ then $\xi_{tr}=f_{tr}$ and $\ast \xi_{\theta\phi}= \ast f_{\theta\phi}$ in a flat Minkowskian spacetime with signature $(-+++)$. We also know that the metric in spherical coordinates $(t,r,\theta,\phi)$ will be diagonal $(-1,1,r^{2},r^{2}\:\sin^{2}\theta)$. The metric determinant $g$ will satisfy $\sqrt{-g}=r^{2}\:\sin\theta$. For the alternating tensor spherical components section \ref{nullcase} is useful when considering all these elements. In the Coulomb geometry the only non-zero tetrad vector components for the local plane one will be,

\begin{eqnarray}
V_{(1)}^{t} &=& \xi^{tr}\:\xi_{tr}\:A^{t} = \mid \xi_{tr} \mid^{2}\:A_{t} \label{V1tEX}\\
V_{(1)}^{r} &=& \xi^{rt}\:\xi_{rt}\:A^{r} = 0 \label{V1rEX}\\
V_{(2)}^{t} &=& \mid \xi_{tr} \mid \: \xi^{tr} \: A_{r} = 0 \label{V2tEX}\\
V_{(2)}^{r} &=& \mid \xi_{tr} \mid \: \xi^{rt} \: A_{t} = \mid \xi_{tr} \mid\:\xi_{tr}\:A_{t} \ .\label{V2rEX}
\end{eqnarray}

where $Q = -2\mid \xi_{tr} \mid^{2}$. In the Coulomb geometry the only non-zero tetrad vector components for the local plane two will be,

\begin{eqnarray}
V_{(3)}^{\theta} &=& \mid \xi_{tr} \mid \:\ast \xi^{\theta\phi}\:Y_{\phi} \label{V3thetaEX}\\
V_{(3)}^{\phi} &=& \mid \xi_{tr} \mid \:\ast \xi^{\phi\theta}\:Y_{\theta} = 0 \label{V3phiEX}\\
V_{(4)}^{\theta} &=& \ast \xi^{\theta\phi}\:\ast \xi_{\theta\phi}\:Y^{\theta}  = 0 \label{V4thetaEX}\\
V_{(4)}^{\phi} &=& \ast \xi^{\phi\theta}\:\ast \xi_{\phi\theta}\:Y^{\phi} \ .\label{V4phiEX}
\end{eqnarray}

We notice that using the four-dimensional Lorentz flat Minkowski metric tensor in spherical coordinates allows the introduction at every point of four orthonormal vectors. Let them be $k^{\mu}_{t}=(1,0,0,0)$, $k^{\mu}_{r}=(0,1,0,0)$, $k^{\mu}_{\theta}=(0,0,\frac{1}{r},0)$ and $k^{\mu}_{\phi}=(0,0,0,\frac{1}{r\:\sin\theta})$. Since the vector $k^{\mu}_{\phi}$ has non-trivial $\phi$ components, then we can choose the gauge vector $Y^{\alpha}=k^{\alpha}_{\phi}$. This way, the equation components (\ref{V3thetaEX}) and (\ref{V4phiEX}) will not be trivial. Despite the fact that in this Coulomb geometry we cannot choose the gauge vector $Y^{\alpha}$ to be $A^{\alpha}$ simply because the components of $V_{(3)}^{\alpha}$ and $V_{(4)}^{\alpha}$ will all be zero and nor we can choose $Y^{\alpha}$ to be $\ast A^{\alpha}$ simply because $\ast A^{\alpha}$ does not exist in the Coulomb geometry with source, we can choose it to be the vector $Y^{\alpha}=k^{\alpha}_{\phi}$.
It is also possible to choose the following $Y^{\alpha}=k^{\alpha}_{\phi} + A^{\alpha}$ gauge-vector in plane two. We can then always choose another gauge by implementing $Y^{\alpha}=k^{\alpha}_{t} +A^{\alpha} \rightarrow Y^{\alpha}=k^{\alpha}_{t} + A^{\alpha} + \Lambda_{,\beta}\:g^{\alpha\beta}$. It is an electromagnetic potential gauge transformation for a valid choice of gauge-vector and we can study the tetrad eigenvector transformations in the local plane two exactly as in reference \cite{A} and section \ref{intro}. There would be no mathematical change in the analysis structure. The whole point of this section is to highlight that when we have in flat-Minkowski spacetime the Maxwell equations with sources, then we have to be careful with our choice for the gauge vector $Y^{\alpha}$. With this observation all the analysis about tetrad vector transformations in the local plane two will follow the same lines as in manuscript \cite{A} and the tetrad study originally made in section \ref{nullcase} for Einstein-Maxwell curved spacetimes without sources will stand. Let us remember that for Einstein-Maxwell curved spacetimes these steps were not necessary since there was a natural non-trivial choice $Y^{\alpha}=\ast A^{\alpha}$. In Einstein-Maxwell curved spacetimes with sources we would also have an analogous choice to the one given in this section depending on the case.
\section{Appendix II}
\label{appendixII}

In this section we will show through the Reissner-Nordstr\"{o}m case how the Kernel of the new internal-spacetime mapping satisfies several properties of homomorphisms in a highly non-trivial way. This example is important to clarify unusual specificities of this mapping that in this rather simple Reissner-Nordstr\"{o}m case can be analyzed with ease. The choice $X^{\alpha}=A^{\alpha}$ and $Y^{\alpha}=\ast A^{\alpha}$ could be illustrated through the Reissner-Nordstr\"{o}m geometry. In this particular geometry, $f_{tr}=\xi_{tr}$ and $\ast f_{\theta\phi}=\ast \xi_{\theta\phi}$, therefore, $A_{t} = e/r$, $A_{r} = 0$, $A_{\theta}=0$, $A_{\phi}=0$ and the only non-zero component for $\ast A_{\alpha}$ will be $\ast A_{\phi} = -q\:\cos\theta$. We proceed to write explicitly the only non-zero components of vectors,

\begin{eqnarray}
U^{\rho} &=& \xi^{\rho\lambda}\:\xi_{\tau\lambda}\:A^{\tau} \:
/ \: (\: \sqrt{-Q/2} \: \sqrt{A_{\mu} \ \xi^{\mu\sigma} \
\xi_{\nu\sigma} \ A^{\nu}}\:) \label{U}\\
V^{\rho} &=& \xi^{\rho\lambda}\:A_{\lambda} \:
/ \: (\:\sqrt{A_{\mu} \ \xi^{\mu\sigma} \
\xi_{\nu\sigma} \ A^{\nu}}\:) \label{V}\\
Z^{\rho} &=& \ast \xi^{\rho\lambda} \: \ast A_{\lambda} \:
/ \: (\:\sqrt{\ast A_{\mu}  \ast \xi^{\mu\sigma}
\ast \xi_{\nu\sigma}  \ast A^{\nu}}\:)
\label{Z}\\
W^{\rho} &=& \ast \xi^{\rho\lambda}\: \ast \xi_{\tau\lambda}
\:\ast A^{\tau} \: / \: (\:\sqrt{-Q/2} \: \sqrt{\ast A_{\mu}
\ast \xi^{\mu\sigma} \ast \xi_{\nu\sigma} \ast A^{\nu}}\:) \ .
\label{W}
\end{eqnarray}

that is, the normalized version of (\ref{V1}-\ref{V4}) which will be \cite{monopole},

\begin{eqnarray}
U^{t} &=& - (\sqrt{q^{2}}/q) / \sqrt{1 - {2m \over r} + {q^{2} \over r^{2}}} \label{Ut}\\
V^{r} &=& \sqrt{1 - {2m \over r} + {q^{2} \over r^{2}}} \label{Vr}\\
Z^{\theta} &=& - \sqrt{\cos^{2}\theta} / (r\:\cos\theta)  \label{Ztheta}\\
W^{\phi} &=& -\sqrt{q^{2}}\:\sqrt{\cos^{2}\theta} / (q\:r\:\sin\theta\:\cos\theta) \ .\label{Wphi}
\end{eqnarray}

In this particular coordinate system we would have to be careful because both vectors $V_{(3)}^{\alpha}$ and $V_{(4)}^{\alpha}$ before normalizing would be zero at the coordinate value $\theta = \pi / 2$. As the purpose of this section is not to find suitable coordinate coverings but to show certain group transformation properties for the new mapping on the local plane one, we do not pay attention to this situation for the moment. Using equation (103) from reference \cite{A} we can write,

\begin{eqnarray}
(-Q/2)\:\Lambda^{\alpha} = -C \: V_{(1)}^{\alpha} - D \: V_{(2)}^{\alpha}
+ M \: V_{(3)}^{\alpha} + N \: V_{(4)}^{\alpha} \ , \label{Lambdainv}
\end{eqnarray}

The coefficients $C$ and $D$ are given in equations (\ref{COEFFCAT1RN}-\ref{COEFFDAT2RN}) while the coefficients $M$ and $N$ are given by equations \cite{A},

\begin{eqnarray}
M&=&(-Q/2)\:V_{(3)\sigma}\:\ast \Lambda^{\sigma} / (\:V_{(4)\beta}\:
V_{(4)}^{\beta}\:)\label{COEFFM}\\
N&=&(-Q/2)\:V_{(4)\sigma}\:\ast \Lambda^{\sigma} / (\:V_{(3)\beta}\:
V_{(3)}^{\beta}\:)\ .\label{COEFFN}
\end{eqnarray}

The gradient $\ast \Lambda^{\sigma}$ is the gradient of a different local scalar $\ast \Lambda$ that is used for independent local gauge transformations on the local plane two. The formula (\ref{Lambdainv}) is useful in general, not simply for the Reissner-Nordstr\"{o}m geometry. Next, we consider the transformation from $A_{t} = e/r$ and $A_{r} = 0$ into a new gauge by using a local gauge transformation given by $\Lambda^{t}=n\:k(t)$ and $\Lambda^{r}=0$ where $k(t)$ is just any smooth function of t, for example $k(t)=\cosh(t)$ and $n$ is a natural number. Let us remind ourselves of the notation $\Lambda^{t}=\Lambda_{,t}\:g^{tt}$. We find that $C=n\:\Lambda^{t}/A^{t}$ and $D=0$. Let us suppose that $1+C>0$ as is the case for the Kernel of the mapping that we are analyzing. This is a case of the kind $C\:\:A^{\mu} = \Lambda^{\mu}$. We know because it has been studied in detail in reference \cite{A} that the inverse local gauge transformation is given by $C^{inv}=-n\:\Lambda^{t}/A^{t}$ and $D^{inv}=0$. We will also consider that in a region of spacetime $C>0$ while $C^{inv}<0$. We will also consider that $1+C>0$ while $1+C^{inv}<0$ for n large enough. From equations (\ref{TN1N}-\ref{TN2N}) we notice the following. For the tetrad gauge transformation induced by $C=n\:\Lambda^{t}/A^{t}$ and $D=0$ we obtain the identity since in this case $|1+C|=1+C$. For the tetrad gauge transformation induced by $C^{inv}=-n\:\Lambda^{t}/A^{t}$ and $D^{inv}=0$ we obtain minus the identity since in this case $1+C^{inv}=1-C<0$ for n large enough. Then we wonder how it is possible that for a tetrad local gauge transformation we obtain the identity and for the inverse we obtain minus the identity. This fact would be at first sight contrary to the group law that we already know this mapping satisfies from our work in a future reference. Let us the see the details and we will prove that the results are consistent.

\subsection{Direct test of the group law}
\label{dtestgl}

According to equations (54-55) in reference \cite{A} section ``gauge geometry'', the coefficients of a local electromagnetic gauge transformation $\Lambda$ of the two vectors that span the local plane one are given by,

\begin{eqnarray}
C&=&(-Q/2)\:V_{(1)\sigma}\:\Lambda^{\sigma} / (\:V_{(2)\beta}\:
V_{(2)}^{\beta}\:)\label{COEFFC}\\
D&=&(-Q/2)\:V_{(2)\sigma}\:\Lambda^{\sigma} / (\:V_{(1)\beta}\:
V_{(1)}^{\beta}\:)\ .\label{COEFFD}
\end{eqnarray}

We would like to calculate the norm of the transformed vectors $\tilde{V}_{(1)}^{\alpha}$ and $\tilde{V}_{(2)}^{\alpha}$ that according to equations (56-57) in reference \cite{A} are given by,

\begin{eqnarray}
\tilde{V}_{(1)}^{\alpha}\:\tilde{V}_{(1)\alpha} &=&
[(1+C)^2-D^2]\:V_{(1)}^{\alpha}\:V_{(1)\alpha}\label{FP}\\
\tilde{V}_{(2)}^{\alpha}\:\tilde{V}_{(2)\alpha} &=&
[(1+C)^2-D^2]\:V_{(2)}^{\alpha}\:V_{(2)\alpha}\ ,\label{SP}
\end{eqnarray}

where the relation $V_{(1)}^{\alpha}\:V_{(1)\alpha} = -V_{(2)}^{\alpha}\:V_{(2)\alpha}$ has been used. Finally we find the transformation of the two vectors spanning the local plane one as in equations (58-59) in reference \cite{A},

\begin{eqnarray}
{\tilde{V}_{(1)}^{\alpha}
\over \sqrt{-\tilde{V}_{(1)}^{\beta}\:\tilde{V}_{(1)\beta}}}&=&
{(1+C) \over \sqrt{(1+C)^2-D^2}}
\:{V_{(1)}^{\alpha} \over \sqrt{-V_{(1)}^{\beta}\:V_{(1)\beta}}}+
{D \over \sqrt{(1+C)^2-D^2}}
\:{V_{(2)}^{\alpha} \over \sqrt{V_{(2)}^{\beta}\:V_{(2)\beta}}}\label{TN1}\\
{\tilde{V}_{(2)}^{\alpha}
\over \sqrt{\tilde{V}_{(2)}^{\beta}\:\tilde{V}_{(2)\beta}}}&=&
{D \over \sqrt{(1+C)^2-D^2}}
\:{V_{(1)}^{\alpha} \over \sqrt{-V_{(1)}^{\beta}\:V_{(1)\beta}}} +
{(1+C) \over \sqrt{(1+C)^2-D^2}}
\:{V_{(2)}^{\alpha} \over \sqrt{V_{(2)}^{\beta}\:V_{(2)\beta}}}\ .
\label{TN2}
\end{eqnarray}

The condition $[(1+C)^2-D^2]>0$ allows for two possible situations, $1+C > 0$ or $1+C < 0$. For the particular case when $1+C > 0$, the transformations
(\ref{TN1}-\ref{TN2}) are telling us that an electromagnetic gauge transformation on the vector field $A^{\alpha}$, that leaves invariant the electromagnetic field $f_{\mu\nu}$, generates a boost transformation on the normalized tetrad vector fields $\left({V_{(1)}^{\alpha} \over \sqrt{-V_{(1)}^{\beta}\:V_{(1)\beta}}}, {V_{(2)}^{\alpha} \over \sqrt{V_{(2)}^{\beta}\:V_{(2)\beta}}}\right)$. Let us consider two local consecutive electromagnetic gauge transformations $\Lambda_{1}$ and $\Lambda_{2}$ that satisfy $1+C_{1} > 0$, $1+C_{2} < 0$ and the proper condition $[(1+C)^2-D^2]>0$ for both. Let us consider the transformation of the coefficients of the first local gauge electromagnetic transformation $\Lambda_{1}$ under the action of the second transformation $\Lambda_{2}$ using always the local vector tetrad transformations (\ref{TN1}-\ref{TN2}) and the norm of the transformed vectors (\ref{FP}-\ref{SP}),

\begin{eqnarray}
\widetilde{C}_{1} &=& (-Q/2)\:\widetilde{V}_{(1)\sigma}\:\Lambda^{\sigma}_{1} / (\:\widetilde{V}_{(2)\beta}\:
\widetilde{V}_{(2)}^{\beta}\:) = \frac{(-Q/2)}{[(1+C_{2})^2-D_{2}^2]}\: \nonumber \\ &&\times\left((1+C_{2})\:{V_{(1)\sigma}\:\Lambda^{\sigma}_{1} \over (V_{(2)}^{\beta}\:V_{(2)\beta})}+D_{2}\:{V_{(2)\sigma}\:\Lambda^{\sigma}_{1} \over (-V_{(1)}^{\beta}\:V_{(1)\beta})}\right)\label{COEFFC2}\\
\widetilde{D}_{1} &=& (-Q/2)\:\widetilde{V}_{(2)\sigma}\:\Lambda^{\sigma}_{1} / (\:\widetilde{V}_{(1)\beta}\:
\widetilde{V}_{(1)}^{\beta}\:) = (-)\frac{(-Q/2)}{[(1+C_{2})^2-D_{2}^2]}\: \nonumber \\ &&\times\left(D_{2}
\:{V_{(1)\sigma}\:\Lambda^{\sigma}_{1} \over (V_{(2)}^{\beta}\:V_{(2)\beta})} +
(1+C_{2})\:{V_{(2)\sigma}\:\Lambda^{\sigma}_{1} \over (-V_{(1)}^{\beta}\:V_{(1)\beta})}\right)\ .\label{COEFFD2}
\end{eqnarray}

After some calculations we finally obtain,

\begin{eqnarray}
\widetilde{C}_{1} &=&  \frac{[(1+C_{2})\:C_{1}-D_{2}\:D_{1}]}{[(1+C_{2})^2-D_{2}^2]} \label{COEFFCF}\\ \nonumber \\
\widetilde{D}_{1} &=&  \frac{[(1+C_{2})\:D_{1}-D_{2}\:C_{1}]}{[(1+C_{2})^2-D_{2}^2]}\ .\label{COEFFDF}
\end{eqnarray}

It is a matter of some algebra using equations (\ref{COEFFCF}-\ref{COEFFDF}) to find the equality,

\begin{eqnarray}
(1 + \widetilde{C}_{1})^{2} - \widetilde{D}_{1}^{2} =  \frac{[(1+C_{1}+C_{2})^{2} - (D_{1}+D_{2})^{2}]}{[(1+C_{2})^2-D_{2}^2]} \ . \label{COEFFGT}
\end{eqnarray}

Next we proceed to write the whole complete local sequence of electromagnetic gauge transformations $\Lambda_{1}$ and $\Lambda_{2}$ of the vectors that span the local plane one,

\begin{eqnarray}
{\widetilde{\widetilde{V}}_{(1)}^{\alpha}
\over \sqrt{-\widetilde{\widetilde{V}}_{(1)}^{\beta}\:\widetilde{\widetilde{V}}_{(1)\beta}}}&=&
{(1+\widetilde{C}_{1}) \over \sqrt{(1+\widetilde{C}_{1})^2-\widetilde{D}_{1}^2}}
\:{\widetilde{V}_{(1)}^{\alpha} \over \sqrt{-\widetilde{V}_{(1)}^{\beta}\:\widetilde{V}_{(1)\beta}}}+
{\widetilde{D}_{1} \over \sqrt{(1+\widetilde{C}_{1})^2-\widetilde{D}_{1}^2}}
\:{\widetilde{V}_{(2)}^{\alpha} \over \sqrt{\widetilde{V}_{(2)}^{\beta}\:\widetilde{V}_{(2)\beta}}}\label{TN1S}\\
{\widetilde{\widetilde{V}}_{(2)}^{\alpha}
\over \sqrt{\widetilde{\widetilde{V}}_{(2)}^{\beta}\:\widetilde{\widetilde{V}}_{(2)\beta}}}&=&
{\widetilde{D}_{1} \over \sqrt{(1+\widetilde{C}_{1})^2-\widetilde{D}_{1}^2}}
\:{\widetilde{V}_{(1)}^{\alpha} \over \sqrt{-\widetilde{V}_{(1)}^{\beta}\:\widetilde{V}_{(1)\beta}}} +
{(1+\widetilde{C}_{1}) \over \sqrt{(1+\widetilde{C}_{1})^2-\widetilde{D}_{1}^2}}
\:{\widetilde{V}_{(2)}^{\alpha} \over \sqrt{\widetilde{V}_{(2)}^{\beta}\:\widetilde{V}_{(2)\beta}}} \ ,
\label{TN2S}
\end{eqnarray}

where,

\begin{eqnarray}
{\tilde{V}_{(1)}^{\alpha}
\over \sqrt{-\tilde{V}_{(1)}^{\beta}\:\tilde{V}_{(1)\beta}}}&=&
{(1+C_{2}) \over \sqrt{(1+C_{2})^2-D_{2}^2}}
\:{V_{(1)}^{\alpha} \over \sqrt{-V_{(1)}^{\beta}\:V_{(1)\beta}}}+
{D_{2} \over \sqrt{(1+C_{2})^2-D_{2}^2}}
\:{V_{(2)}^{\alpha} \over \sqrt{V_{(2)}^{\beta}\:V_{(2)\beta}}}\label{TN1T}\\
{\tilde{V}_{(2)}^{\alpha}
\over \sqrt{\tilde{V}_{(2)}^{\beta}\:\tilde{V}_{(2)\beta}}}&=&
{D_{2} \over \sqrt{(1+C_{2})^2-D_{2}^2}}
\:{V_{(1)}^{\alpha} \over \sqrt{-V_{(1)}^{\beta}\:V_{(1)\beta}}} +
{(1+C_{2}) \over \sqrt{(1+C_{2})^2-D_{2}^2}}
\:{V_{(2)}^{\alpha} \over \sqrt{V_{(2)}^{\beta}\:V_{(2)\beta}}}\ .
\label{TN2T}
\end{eqnarray}

After we make use of equations (\ref{COEFFCF}-\ref{COEFFGT}) we can write after some algebraic calculations the final expressions for the sequence of two local electromagnetic gauge transformations $\Lambda_{1}$ and $\Lambda_{2}$ of the two vectors that span the local plane one,

\begin{eqnarray}
{\widetilde{\widetilde{V}}_{(1)}^{\alpha}
\over \sqrt{-\widetilde{\widetilde{V}}_{(1)}^{\beta}\:\widetilde{\widetilde{V}}_{(1)\beta}}}&=&
{[1+C_{1}+C_{2}] \over \sqrt{(1+C_{1}+C_{2})^2-(D_{1}+D_{2})^2}}
\:{V_{(1)}^{\alpha} \over \sqrt{-V_{(1)}^{\beta}\:V_{(1)\beta}}}+ \nonumber \\
&&{[D_{1}+D_{2}] \over \sqrt{(1+C_{1}+C_{2})^2-(D_{1}+D_{2})^2}}
\:{V_{(2)}^{\alpha} \over \sqrt{V_{(2)}^{\beta}\:V_{(2)\beta}}} \label{TN1FINAL}\\
{\widetilde{\widetilde{V}}_{(2)}^{\alpha}
\over \sqrt{\widetilde{\widetilde{V}}_{(2)}^{\beta}\:\widetilde{\widetilde{V}}_{(2)\beta}}}&=&
{[D_{1}+D_{2}] \over \sqrt{(1+C_{1}+C_{2})^2-(D_{1}+D_{2})^2}}
\:{V_{(1)}^{\alpha} \over \sqrt{-V_{(1)}^{\beta}\:V_{(1)\beta}}} + \nonumber \\
&&{[1+C_{1}+C_{2}] \over \sqrt{(1+C_{1}+C_{2})^2-(D_{1}+D_{2})^2}}
\:{V_{(2)}^{\alpha} \over \sqrt{V_{(2)}^{\beta}\:V_{(2)\beta}}} \ .
\label{TN2FINAL}
\end{eqnarray}

As we can see the local transformation of the two vectors that span the local plane one by two consecutive local electromagnetic gauge transformations that satisfy the proper condition $[(1+C)^2-D^2]>0$, is found to be equal to the addition of both as it was expected from an Abelian group even though highly not-trivial because of the relations (\ref{COEFFCF}-\ref{COEFFDF}).

Let us next consider for our particular example that $C_{1}=C>0$, $D_{1}=0$, $C_{2}=C^{inv}=-C_{1}=-C<0$ and $D_{2}=0$. We also know from our assumptions in the beginning of section \ref{appendixII} that for $n$ large enough $1+C_{1}=1+C>0$ and $1+C_{2}=1-C<0$. If we use equations (\ref{COEFFCF}-\ref{COEFFDF}) we find,

\begin{eqnarray}
\widetilde{C}_{1} &=&  \frac{[(1+C_{2})\:C_{1}-D_{2}\:D_{1}]}{[(1+C_{2})^2-D_{2}^2]} = \frac{C_{1}}{1-C_{1}} \label{COEFFCFRNE}\\ \nonumber \\
\widetilde{D}_{1} &=&  \frac{[(1+C_{2})\:D_{1}-D_{2}\:C_{1}]}{[(1+C_{2})^2-D_{2}^2]} = 0 \ .\label{COEFFDFRNE}
\end{eqnarray}

We also find

\begin{eqnarray}
(1 + \widetilde{C}_{1})^{2} - \widetilde{D}_{1}^{2} =  \frac{[(1+C_{1}+C_{2})^{2} - (D_{1}+D_{2})^{2}]}{[(1+C_{2})^2-D_{2}^2]} = \frac{1}{(1-C_{1})^{2}}\ . \label{COEFFGTRNE}
\end{eqnarray}

When we use in sequence equations (\ref{TN1S}-\ref{TN2S}) and (\ref{TN1T}-\ref{TN2T}) we find,

\begin{eqnarray}
{\widetilde{\widetilde{V}}_{(1)}^{\alpha}
\over \sqrt{-\widetilde{\widetilde{V}}_{(1)}^{\beta}\:\widetilde{\widetilde{V}}_{(1)\beta}}}&=&
{|1-C_{1}| \over (1-C_{1})}
\:{\widetilde{V}_{(1)}^{\alpha} \over \sqrt{-\widetilde{V}_{(1)}^{\beta}\:\widetilde{V}_{(1)\beta}}} \label{TN1SRNE}\\
{\widetilde{\widetilde{V}}_{(2)}^{\alpha}
\over \sqrt{\widetilde{\widetilde{V}}_{(2)}^{\beta}\:\widetilde{\widetilde{V}}_{(2)\beta}}}&=&
{|1-C_{1}| \over (1-C_{1})}
\:{\widetilde{V}_{(2)}^{\alpha} \over \sqrt{\widetilde{V}_{(2)}^{\beta}\:\widetilde{V}_{(2)\beta}}} \ ,
\label{TN2SRNE}
\end{eqnarray}

where,

\begin{eqnarray}
{\tilde{V}_{(1)}^{\alpha}
\over \sqrt{-\tilde{V}_{(1)}^{\beta}\:\tilde{V}_{(1)\beta}}}&=&
{(1-C_{1}) \over |1-C_{1}|}
\:{V_{(1)}^{\alpha} \over \sqrt{-V_{(1)}^{\beta}\:V_{(1)\beta}}} \label{TN1TRNE}\\
{\tilde{V}_{(2)}^{\alpha}
\over \sqrt{\tilde{V}_{(2)}^{\beta}\:\tilde{V}_{(2)\beta}}}&=&
{(1-C_{1}) \over |1-C_{1}|}
\:{V_{(2)}^{\alpha} \over \sqrt{V_{(2)}^{\beta}\:V_{(2)\beta}}}\ .
\label{TN2TRNE}
\end{eqnarray}

We have proved in the end that the apparent contradiction of having one transformation with $1+C>0$ to be the identity while the inverse with $1+C^{inv}=1-C<0$ is minus the identity as far as it regards the group law is not a true inconsistence because the composition transformation is not just the product of these two matrices giving minus the identity but a far more involved and non-trivial composition as presented through the whole work in this section. From equations (\ref{TN1SRNE}-\ref{TN2SRNE}) and (\ref{TN1TRNE}-\ref{TN2TRNE}) we have proved that the composition is consistent with the direct result obtained from equations (\ref{TN1FINAL}-\ref{TN2FINAL}) where $1+C_{1}+C_{2}=1$ and $D_{1}+D_{2}=0$. That is, the identity in both separate calculations.

\subsection{Reverse test of the group law}
\label{dtestgl}

Let us next consider the reverse order in gauge tetrad transformations for our particular example such that $C_{1}=-C<0$, $D_{1}=0$, $C_{2}=-C^{inv}=-C_{1}=C>0$ and $D_{2}=0$. We also know from our assumptions in the beginning of section \ref{appendixII} that for $n$ large enough $1+C_{1}=1-C<0$ and $1+C_{2}=1+C>0$. If we use equations (\ref{COEFFCF}-\ref{COEFFDF}) we find,

\begin{eqnarray}
\widetilde{C}_{1} &=&  \frac{[(1+C_{2})\:C_{1}-D_{2}\:D_{1}]}{[(1+C_{2})^2-D_{2}^2]} = \frac{C_{1}}{1-C_{1}} \label{COEFFCFRNERO}\\ \nonumber \\
\widetilde{D}_{1} &=&  \frac{[(1+C_{2})\:D_{1}-D_{2}\:C_{1}]}{[(1+C_{2})^2-D_{2}^2]} = 0 \ .\label{COEFFDFRNERO}
\end{eqnarray}

We also find

\begin{eqnarray}
(1 + \widetilde{C}_{1})^{2} - \widetilde{D}_{1}^{2} =  \frac{[(1+C_{1}+C_{2})^{2} - (D_{1}+D_{2})^{2}]}{[(1+C_{2})^2-D_{2}^2]} = \frac{1}{(1-C_{1})^{2}}\ . \label{COEFFGTRNERO}
\end{eqnarray}

When we use in sequence equations (\ref{TN1S}-\ref{TN2S}) and (\ref{TN1T}-\ref{TN2T}) we find,

\begin{eqnarray}
{\widetilde{\widetilde{V}}_{(1)}^{\alpha}
\over \sqrt{-\widetilde{\widetilde{V}}_{(1)}^{\beta}\:\widetilde{\widetilde{V}}_{(1)\beta}}}&=&
{|1-C_{1}| \over (1-C_{1})}
\:{\widetilde{V}_{(1)}^{\alpha} \over \sqrt{-\widetilde{V}_{(1)}^{\beta}\:\widetilde{V}_{(1)\beta}}} \label{TN1SRNERO}\\
{\widetilde{\widetilde{V}}_{(2)}^{\alpha}
\over \sqrt{\widetilde{\widetilde{V}}_{(2)}^{\beta}\:\widetilde{\widetilde{V}}_{(2)\beta}}}&=&
{|1-C_{1}| \over (1-C_{1})}
\:{\widetilde{V}_{(2)}^{\alpha} \over \sqrt{\widetilde{V}_{(2)}^{\beta}\:\widetilde{V}_{(2)\beta}}} \ ,
\label{TN2SRNERO}
\end{eqnarray}

where,

\begin{eqnarray}
{\tilde{V}_{(1)}^{\alpha}
\over \sqrt{-\tilde{V}_{(1)}^{\beta}\:\tilde{V}_{(1)\beta}}}&=&
{(1-C_{1}) \over |1-C_{1}|}
\:{V_{(1)}^{\alpha} \over \sqrt{-V_{(1)}^{\beta}\:V_{(1)\beta}}} \label{TN1TRNERO}\\
{\tilde{V}_{(2)}^{\alpha}
\over \sqrt{\tilde{V}_{(2)}^{\beta}\:\tilde{V}_{(2)\beta}}}&=&
{(1-C_{1}) \over |1-C_{1}|}
\:{V_{(2)}^{\alpha} \over \sqrt{V_{(2)}^{\beta}\:V_{(2)\beta}}}\ .
\label{TN2TRNERO}
\end{eqnarray}

We notice that the difference between equations (\ref{TN1SRNE}-\ref{TN2SRNE}) plus (\ref{TN1TRNE}-\ref{TN2TRNE}) on one hand and equations (\ref{TN1SRNERO}-\ref{TN2SRNERO}) plus (\ref{TN1TRNERO}-\ref{TN2TRNERO}) on the other hand is that in the first case ${|1-C_{1}| \over (1-C_{1})}=-1$ while in the second case ${|1-C_{1}| \over (1-C_{1})}=+1$. We have proved in the end that the apparent contradiction of having one transformation with $1-C<0$ to be minus the identity while the inverse with $1-C^{inv}=1+C>0$ to be the identity as far as it regards the group law is not a true inconsistence because the composition transformation is not just the product of these two matrices giving minus the identity but a far more involved and non-trivial composition as presented through the whole work in this section. From equations (\ref{TN1SRNERO}-\ref{TN2SRNERO}) and (\ref{TN1TRNERO}-\ref{TN2TRNERO}) we have proved that the composition is consistent with the direct result obtained from equations (\ref{TN1FINAL}-\ref{TN2FINAL}) where $1+C_{1}+C_{2}=1$ and $D_{1}+D_{2}=0$. That is, the identity in both separate calculations also for the group law with the transformations in reverse order.


\end{document}